\documentclass[sigconf]{acmart}
\usepackage{dirtytalk}
\usepackage{multirow}
\usepackage{colortbl}
\usepackage{graphicx}
\usepackage{subcaption}
\usepackage{multicol}
\usepackage{multirow}
\usepackage{pifont}

\usepackage{tcolorbox}
\usepackage{listings}
\usepackage{xcolor}  
\tcbuselibrary{listingsutf8}

\usepackage[export]{adjustbox}

\setcopyright{acmlicensed}
\copyrightyear{2024}
\acmYear{2024}
\acmDOI{XXXXXXX.XXXXXXX}

\acmConference[]{}{June 03--05,
  2024}{}
\acmISBN{978-1-4503-XXXX-X/18/06}

\begin{document}

\title{Search Engines in an AI Era: The False Promise of Factual and Verifiable Source-Cited Responses}


\author{Pranav Narayanan Venkit}
\email{pranav.venkit@psu.edu}
\affiliation{%
  \institution{Pennsylvania State University}
  \city{University Park}
  \state{Pennsylvania}
  \country{USA}
}
\author{Philippe Laban}
\email{plaban@salesforce.com}
\affiliation{%
  \institution{Salesforce AI Research}
  \city{Palo Alto}
  \state{California}
  \country{USA}
}
\author{Yilun Zhou}
\email{yilun.zhou@salesforce.com}
\affiliation{%
  \institution{Salesforce AI Research}
  \city{Palo Alto}
  \state{California}
  \country{USA}
}
\author{Yixin Mao}
\email{y.mao@salesforce.com}
\affiliation{%
  \institution{Salesforce AI Research}
  \city{Palo Alto}
  \state{California}
  \country{USA}
}
\author{Chien-Sheng Wu}
\email{wu.jason@salesforce.com}
\affiliation{%
  \institution{Salesforce AI Research}
  \city{Palo Alto}
  \state{California}
  \country{USA}
}
\renewcommand{\shortauthors}{Trovato et al.}

\newcommand{\yilun}[1]{\textcolor{blue}{(Yilun: #1)}}
\newcommand{\yixin}[1]{\textcolor{orange}{(Yixin: #1)}}
\newcommand{\jw}[1]{\textcolor{cyan}{[Jason: #1]}}

\definecolor{colorans}{HTML}{E61670}
\definecolor{colorcit}{HTML}{4A86E8}
\definecolor{colorsrc}{HTML}{49C39E}
\definecolor{colorint}{HTML}{5900A7}

\definecolor{redlight}{HTML}{ef2e65}
\definecolor{orangelight}{HTML}{edbf2f}
\definecolor{greenlight}{HTML}{16c056}

\newcommand{\redcircle}{\textcolor{redlight}{\ding{116}}}
\newcommand{\orangecircle}{\textcolor{orangelight}{\ding{108}}}
\newcommand{\greencircle}{\textcolor{greenlight}{\ding{115}}}

\newcommand{\symbolimg}[2][0.4cm]{%
  \includegraphics[height=#1,valign=c]{#2}%
}

\lstset{basicstyle=\ttfamily,breaklines=true}

\begin{abstract}
Large Language Model (LLM)-based applications are graduating from research prototypes to products serving millions of users, influencing how people write and consume information. A prominent example is the appearance of Answer Engines: LLM-based generative search engines supplanting traditional search engines. Answer engines not only retrieve relevant sources to a user query but synthesize answer summaries that cite the sources. To understand these systems' limitations, we first conducted a study with 21 participants, evaluating interactions with answer vs. traditional search engines and identifying 16 answer engine limitations. From these insights, we propose 16 answer engine design recommendations, linked to 8 metrics. An automated evaluation implementing our metrics on three popular engines (You.com, Perplexity.ai, BingChat) quantifies common limitations (e.g., frequent hallucination, inaccurate citation) and unique features (e.g., variation in answer confidence), with results mirroring user study insights. We release our Answer Engine Evaluation benchmark (AEE) to facilitate transparent evaluation of LLM-based applications.

\end{abstract}

\ccsdesc[500]{Human-centered computing~HCI design and evaluation methods}
\ccsdesc[500]{Human-centered computing~User studies}
\ccsdesc[500]{Human-centered computing~Empirical studies in HCI}
\ccsdesc[500]{Human-centered computing~Natural language interfaces}
\ccsdesc[500]{Computing methodologies~Natural language generation}
\ccsdesc[500]{Information systems~Information retrieval}

\keywords{Answer Engines, Generative Search Engine, RAG systems, Ethical Audit, User Study Evaluation, Fairness and Ethics}



\begin{teaserfigure}
    \centering
    \includegraphics[width=0.9\textwidth]{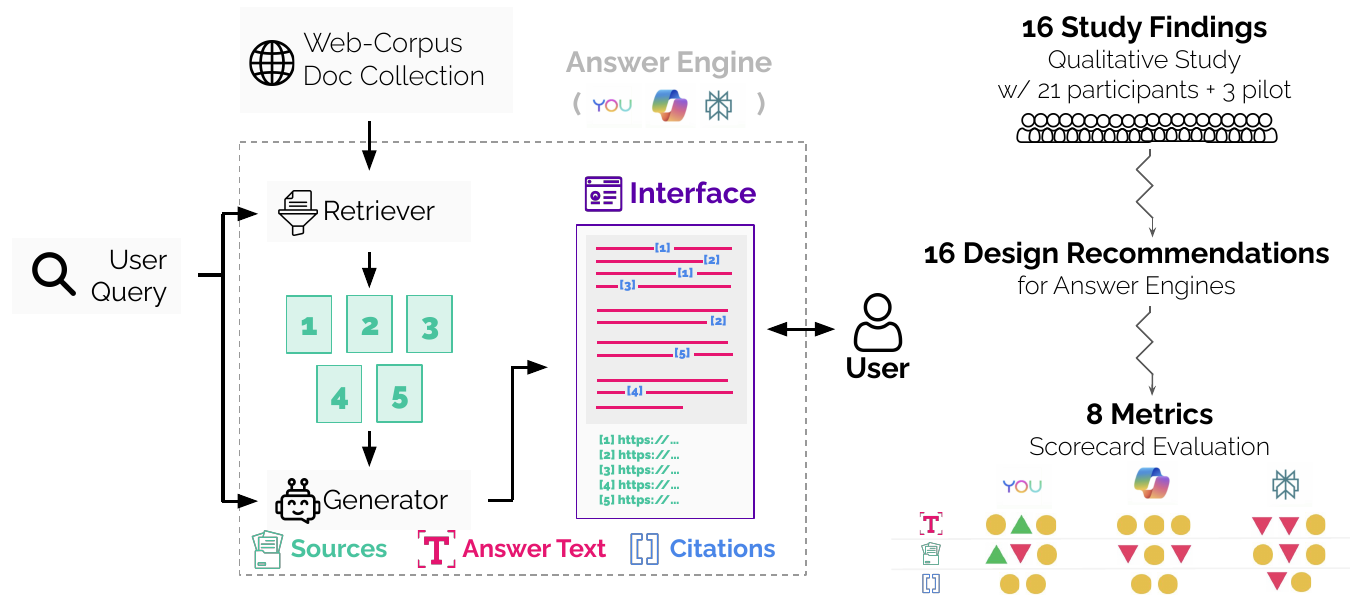}
    \caption{The design pipeline of an answer engine and the study framework used to audit these systems. The figure illustrates the key components of an answer engine, including how it generates answers based on user queries, with a focus on outputs such as \textbf{\textcolor{colorsrc}{\symbolimg{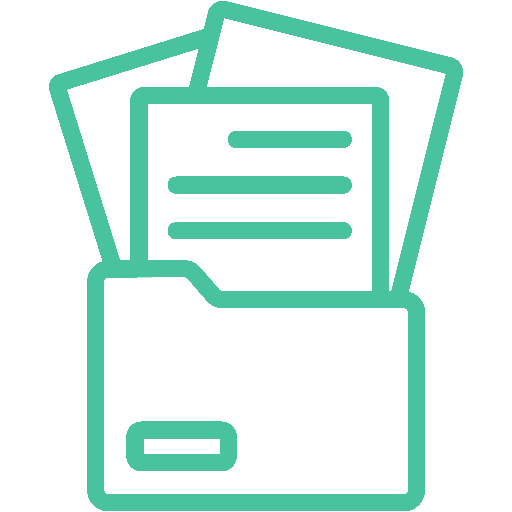} sources}},  \textbf{\textcolor{colorans}{\symbolimg{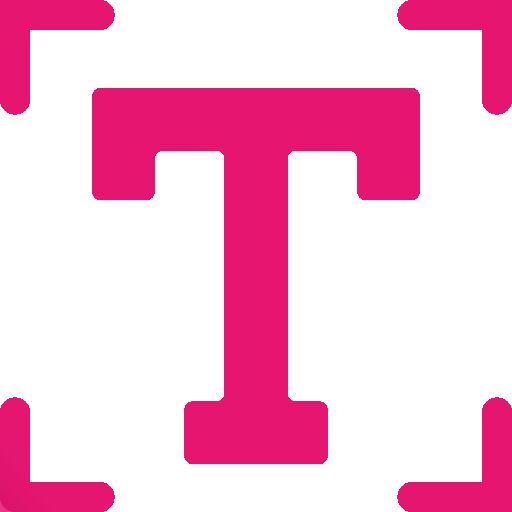} answer text}}, and \textbf{\textcolor{colorcit}{\symbolimg{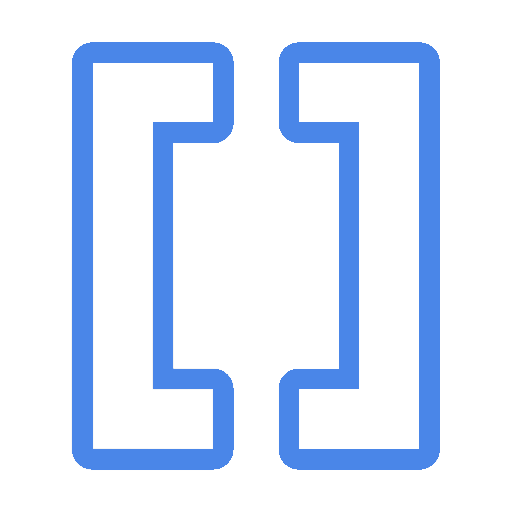} citations}}. On the right, a summary of the findings from our usability study is presented with the final scorecard evaluation of You Chat, Bing Copilot and Perplexity accordingly.}
    \label{fig:paper_teaser}
\end{teaserfigure}
\maketitle

\section{Introduction}
Large Language Models have recently become part of daily life for many, with services such as ChatGPT and Claude offering AI-based conversational assistance to hundreds of millions of customers \cite{narayanan2023towards, ferrara2024genai, pulapaka2024genai}. In doing so, such systems have graduated from academic tools that were evaluated from a technical standpoint to \textit{sociotechnical systems} \cite{cooper1971sociotechnical} that have both technical and social impact, and that require more nuanced evaluation, as they can influence various facets of society, including communication, information dissemination, and decision-making \cite{shah2024envisioning, narayanan2023towards}.

A prominent example of an LLM-based sociotechnical system is the \textit{Answer Engine}, also known as a Generative Search Engine, illustrated in Figure~\ref{fig:paper_teaser}. Answer Engines are marketed as replacements for traditional search engines -- such as Google or Bing -- and work in the following retrieval-augmented way: a user with an information need formulates a search query \cite{memon2024search}. The Answer Engine first retrieves relevant \textcolor{colorsrc}{\symbolimg{Images/icons/sources_color} source documents} that likely contain answer elements to the user's query, leveraging a retrieval system (which can be a traditional search engine). The Answer Engine then composes a textual prompt that contains the user's query, and the retrieved sources, and instructs an LLM to generate a long and self-contained \textcolor{colorans}{\symbolimg{Images/icons/answer_text_color} answer} for the user, based on the content of the sources. Crucially, \textcolor{colorcit}{\symbolimg{Images/icons/citation_color} citations} are inserted into the answer, with each citation linking to the sources that support each statement within the answer.
This citation-enriched answer is provided to the user in a \textcolor{colorint}{\symbolimg{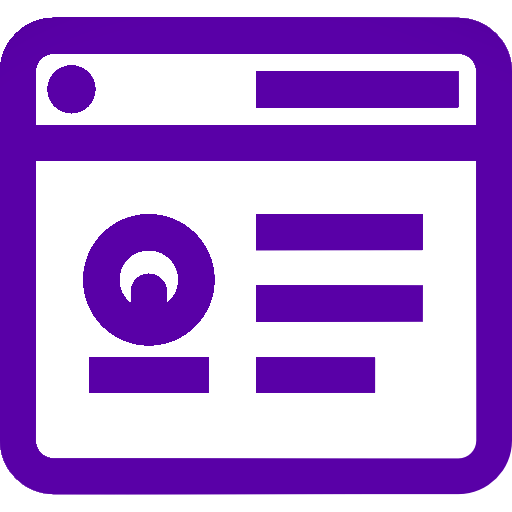} user interface}: the citation forms the semantic glue between the generated answer and the sources, with a click on a citation allowing the user to navigate to the source or sources that support any statement.

In essence, the answer engine promises a streamlining of a user's information-seeking journey \cite{shah2024envisioning}. The Answer Engine concisely summarizes the information the user is looking for, and sources remain within a click in case the user desires to deepen their understanding or verify the information's veracity at the source. Recently, several free Answer Engines have become popular such as You.com, Perplexity.ai, and Bing Chat, with some reporting millions of daily searches performed by their users: Answer Engines are answering user needs.

However, there are several well-known limitations to Answer Engines, primarily stemming from the use of LLMs as part of answer generation. First, LLMs are known to hallucinate information and cannot detect factual inconsistencies \cite{venkit2024confidently, huang2023survey}, even when authoritative sources are provided. Second, prior work \cite{liu2023evaluating,laban2024summary} has also shown limitations of answer engines' ability to assess the accuracy of citations within an answer. Third, LLMs accumulate knowledge within their internal weights during pre-training, and prior work has shown limited success at enforcing that an LLM generates information based solely on documents provided in a prompt rather than based on pretraining information which can be noisy or outdated \cite{kaur-etal-2024-evaluating}. Finally, such systems exhibit sycophantic behavior: a preference for the system to agree with the user's perceived opinion over objective truth \cite{sharma2024generative,laban2023you}. 

All these known limitations lead to a potential impact on the quality of generated answers, negatively affecting user experience. Yet prior work has evaluated LLMs and their output primarily from a technical perspective \cite{es2023ragas, laban2024summary}. Since Answer Engines are used by millions daily, it is equally important to evaluate them from a social perspective, to understand how users perceive Answer Engines, and how they navigate limitations.

We start our work with an audit-centric usability study (Section~\ref{sec:study}) involving 24 participants\footnote{Pilot study with 3 participants and a final usability study with 21 participants} with expertise in technical domains (e.g., sociology, economics). Participants interact with Answer Engines and traditional Search Engines on two kinds of search queries: \textit{expertise} and \textit{debate} queries. Expertise queries are technical queries that participants self-report being experts on. Participants' familiarity with the answer allows us to evaluate how Answer Engines perform on deeply technical questions. Debate queries are queries related to a debate topic, formulated either to be pro or against the debate (e.g., ``Why should we abolish Daylight''). By initially asking participants if they support one side of the debate, we can evaluate how participants interact with the answers that support or refute their opinions.

The usability study follows a think-aloud protocol and enables us to obtain two main kinds of insights: (i) quantitative insights on how users interact with answers, citations, and sources in both Answer Engines and traditional Search Engines, (ii) qualitative feedback from participants which we group using inductive reasoning \cite{glaser1967discovery,glaser1992basics} followed by a qualitative coding method \cite{st2014qualitative, auerbach2003qualitative} into \textbf{16 insights on the limitations of Answer Engines}.
With the study completed, we then propose \textbf{16 Design Recommendations} that are both actionable and measurable, as we design \textbf{8 quantitative metrics} that tie the design recommendations to specific measures (Section~\ref{sec:recs_and_metrics}). 

Finally, we implement a large-scale automated evaluation of three popular Answer Engines (YouChat, Bing Copilot, and Perplexity AI\footnote{You Chat: https://you.com/ ; Copilot: https://www.bing.com/chat ; Perplexity: https://www.perplexity.ai/}) using the 8 metrics, on 303 search queries from our usability study. We consolidate the metrics into a scorecard, the Answer Engine Evaluation (AEE) benchmark, for each Answer Engine. One of our findings show that all evaluated answer engines frequently generate one-sided answers (50-80\%) that favor agreement with charged debate questions, with Perplexity performing the worst, in multiple aspects, despite generating the longest answers, indicating that increasing answer length does not improve answer diversity. We release our automatic evaluation framework publicly, to encourage the community to evaluate Answer Engines as the technology evolves and matures\footnote{https://github.com/SalesforceAIResearch/answer-engine-eval}. 

Through this work, we hence (i) conduct a usability study of answer engines to audit their societal implications, (ii) propose design recommendations linked to automated evaluation metrics, (iii) perform a qualitative analysis of existing public systems to identify issues and create a model performance scorecard, and (iv) explore societal dynamics and user-driven recommendations.

\section{Background and Related Works}

\subsection{Understanding AI of Today}

As AI becomes more embedded in daily life, their role has evolved from simple technical tools to complex sociotechnical systems. These systems involve an intricate interplay between social actors and technological components that together shape goal-oriented behaviors \cite{cooper1971sociotechnical, venkit2023sentiment, narayanan2023towards, venkit2024confidently}. AI systems in domains like education \cite{holmes2022state}, healthcare \cite{arefin2024ai}, and policymaking \cite{capraro2024impact} are thus deeply entwined with the social practices and institutional contexts in which they operate \cite{gautam2024melting}.

However, despite the recognition of AI as inherently sociotechnical, current research often adopts a \textit{technocentric} perspective, focusing on algorithmic and computational aspects while neglecting broader societal implications \cite{narayanan2023towards, dean2021axes, ehsan2020human, venkit2023automated}. This gap is evident in the conceptualization of terms like bias \cite{blodgett2020language}, sentiment \cite{venkit2023sentiment}, and hallucination \cite{venkit2024confidently}. As \citet{venkit2023sentiment} argues, understanding AI through a sociotechnical lens is crucial to fully grasp its impacts, biases, and potential harms. Insights from Human-Centered Explainable AI (HCXAI) emphasize the need for a human-centric approach to technology, focusing on how AI systems can better align with human understanding and accessibility \cite{ehsan2020human, ehsan2024human}. \citet{ehsan2020human} advocates for positioning ``the human'' at the core of technology design, leveraging the social dynamics and context of AI systems to bridge gaps for non-technical users—something that is currently lacking. This approach forms the basis for designing AI systems that are not only transparent but also safer and more trustworthy, addressing a significant gap in AI development today. 

The emergence of answer engines, or RAG (Retrieval Augmented Generation) systems, highlights the need for a social understanding of automated information retrieval. Unlike traditional search engines that provide a ranked list of documents, RAG systems generate synthesized responses by combining a retriever with a generator (often an LLM) to produce answers from these passages \cite{lewis2020pre, fan2024survey, gao2023retrieval, ding2024survey}. While designed to enhance the relevance of retrieved information and address issues like hallucination \cite{rawte2023survey, huang2023survey, venkit2024confidently}, these systems also raise concerns about selective information presentation and bias amplification \cite{shah2024envisioning, lindemann2024chatbots, sharma2024generative, gupta2024sociodemographic}. 



\subsection{The Shortcomings of Answer Engines}

Answer engines are marketed as efficient tools for simplifying information retrieval by reducing the need for users to manually sift through data repositories \cite{Robison2024, Seetharaman2024}. However, recent developments, such as Google AI Overview and Perplexity, have exposed new ethical challenges and negative user experiences. For example, Google's answer engine erroneously advised users to ``put glue on their pizza,'' revealing how the system misinterpreted sarcastic content from the internet, presenting it as fact with undue authority\footnote{https://www.theverge.com/2024/5/23/24162896/google-ai-overview-hallucinations-glue-in-pizza}. Such cases of misinformation highlight the risks associated with automating information retrieval, especially under the guise of `Google doing the Googling’ for users \cite{Robison2024, Seetharaman2024, vogt2024glue}.


The release of OpenAI's `SearchGPT,' marketed as a `Google search killer' \cite{Seetharaman2024}, further exacerbates these concerns. As reliance on these tools grows, so does the urgency to understand their impact. \citet{lindemann2024chatbots} introduces the concept of Sealed Knowledge, which critiques how these systems limit access to diverse answers by condensing search queries into singular, authoritative responses, effectively decontextualizing information and narrowing user perspectives \cite{muhlhoff2018digitale, haraway2013situated}. This ``sealing'' of knowledge perpetuates selection biases and restricts marginalized viewpoints \cite{lindemann2024chatbots}.

Building on this, \citet{sharma2024generative} argues that answer engines foster echo chambers, where exposure to diverse opinions is minimized, reinforcing existing beliefs and reducing the visibility of minority perspectives. This is particularly problematic given the established Western-centric bias in text generation models \cite{ghosh2024generative, narayanan2023unmasking, venkit2023nationality, Naous2023HavingBA}. When integrated into search engines, these models further propagate a predominantly Western viewpoint or an `automated colonial impulse' \cite{das2024colonial, das2022decolonial, das2023studying}, underscoring the need for comprehensive studies on their societal risks.

A key concern surrounding answer engines is their inherently `black-box' nature \cite{o2017weapons}. AI systems are often perceived as black boxes due to the opacity of their decision-making processes and the hidden biases within them, which are obscured by proprietary algorithms and vast training datasets \cite{bender2021dangers, hayes2023generative, venkit2023automated}. Answer engines intensify this problem by merging two opaque systems: a search engine \cite{goldman2005search} and a generative AI model \cite{bender2021dangers}, resulting in compounded opacity and reduced user autonomy.
This dual-layered opacity leads to problematic outcomes, such as those identified by \citet{li2024generative}, who revealed sentiment bias based on query content, along with commercial and geographic biases in the sources answer engines use. The over-reliance on uneven quality sources, heavily skewed toward news, media, and business, further illustrates the need for transparency \cite{li2024generative}. 

\subsection{Beyond a Positivism and Technical Lens of Answer Engines}

As answer engines gain traction within the NLP and AI communities, there has been a notable increase in efforts to evaluate and benchmark their performance \cite{jeong2024adaptive, wu2024faithful, es2023ragas, zhu2024rageval}. However, these benchmarking initiatives have largely maintained a technocentric focus, emphasizing model performance metrics while often overlooking the broader societal implications of these technologies. These behaviours are categorized as positivisms of technology where evaluations are rooted in law-like patterns and cause-effect relationships \cite{wyly2014automated}. We propose the need for a sociotechnical approach to truly understand how these systems can impact society.

For example, the widely used RAGAS benchmark evaluates answer engines using a comprehensive set of metrics, including faithfulness, answer relevance, and context relevance, allowing for assessment across various dimensions without relying on ground truth human annotations \cite{es2024ragas, es2023ragas}. Similarly, the ClashEval framework \cite{wu2024faithful} demonstrates how LLMs can adopt incorrect retrieved content, overriding their own correct prior knowledge more than 60\% of the time. Beyond performance metrics, researchers are actively extending the application of RAG models to socially significant domains \cite{roychowdhury2024evaluation, siriwardhana2023improving}. \citet{siriwardhana2023improving} propose an adaptation of RAG that allows it to integrate domain-specific knowledge bases, enhancing their application in critical fields like medicine and news where accuracy is crucial. Similar adaptations are being explored in telecommunications \cite{roychowdhury2024evaluation}, gaming \cite{chauhan2024evaluating}, and agriculture \cite{gupta2024rag}, reflecting a broader effort to address diverse sociotechnical needs. In this work, we bridge the gap by bringing a sociotechnical perspective to answer engine evaluation, which requires integrating human interactions and societal impacts  \cite{bender2024resisting} into the evaluation process \cite{ehsan2024xai, narayanan2023towards}.



\section{Answer Engine Usability Study} \label{sec:study}


We now describe the usability study we conducted aimed at understanding the societal impact of answer engines when compared to a traditional search engine. We first describe the design of the study (Section~\ref{sec:study_design}), then proceed with the qualitative (Section~\ref{sec:study_qualitative}) and quantitative (Section~\ref{sec:study_quantitative}) findings.

\subsection{Study Design} \label{sec:study_design}

\begin{figure*}[h]
  \centering
  \includegraphics[width=0.85\textwidth]{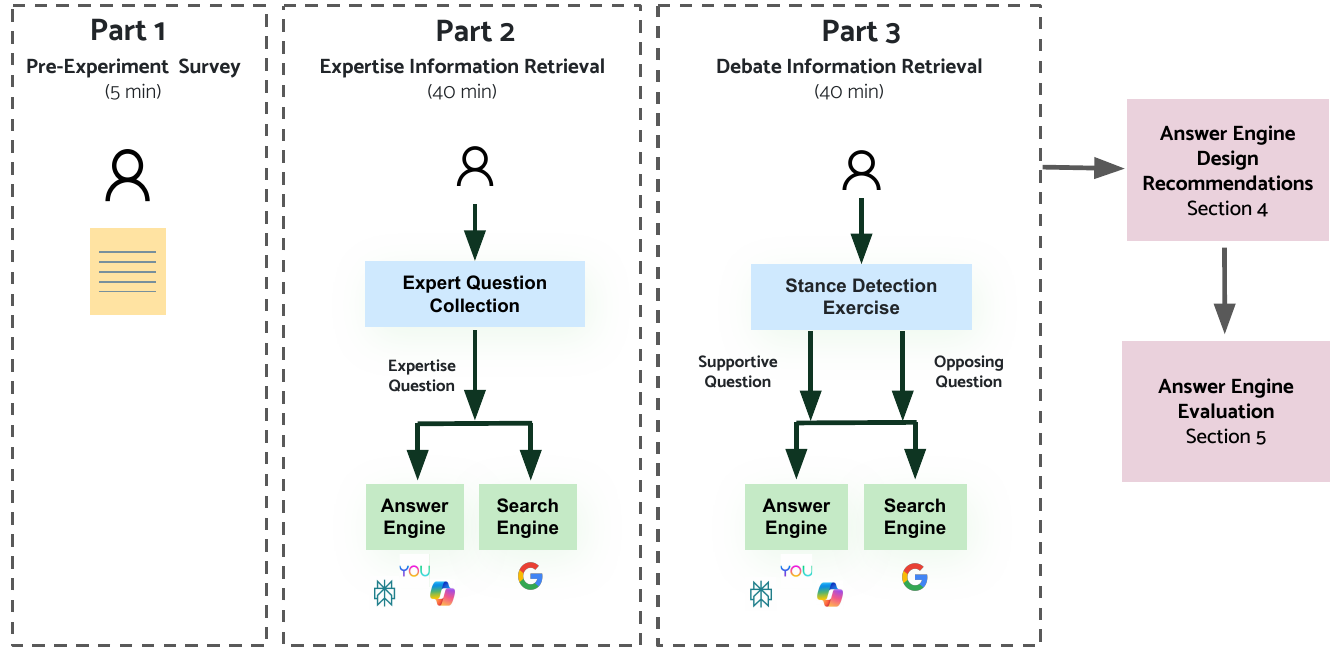}
  \caption{High-level diagram of the three parts to the 90-minute usability study we conducted, and the work that derives from study findings: design recommendations, and the Answer Engine Evaluation (AEE) framework.}
  \label{fig:Qualitative}
\end{figure*}

Figure~\ref{fig:Qualitative} illustrates the study protocol, designed as a 90-minute one-on-one session via video conference, which was recorded and transcribed with participants' consent. We first describe participant recruitment, and then the three steps to the study: \textit{pre-study questionnaire, expert information retrieval,} and \textit{debated information retrieval}.

\subsubsection{Participant Recruitment.} We decided to recruit participants with technical expertise (i.e., having completed or currently pursuing a Ph.D.), as it would allow participants to evaluate the systems on complex queries participants have expertise in. Participant expertise allows us to understand the performance of such systems in realistic but technically advanced topics. Our recruitment criteria targeted experts from diverse academic and professional backgrounds. Participants were recruited through a combination of academic channels (via email invitation and LinkedIn), and social media platforms (via Twitter and Reddit). 
The study was conducted using the \textit{User Interviews platform}\footnote{https://www.userinterviews.com/}, and Google Meet for video-conferencing.

We recruited 24 participants aged 22 to 38 years (M=29.3, SD=2.99), with a gender distribution of 66.67\% female (n=16), 33.34\% male (n=7), and 4.16\% non-binary (n=1). Participants' occupations distribution were 45.83\% PhD Candidates (n=11), 16.67\% research professionals including postdoctoral researchers (n=4), 33.34\% industry experts (n=8), and 4.17\% other professionals (n=1). Regarding participant expertise, 25.00\% were experts in Human-Computer Interaction (n=6), 25.00\% in Artificial Intelligence and Computational Research (n=6), 20.83\% in Healthcare and Medicine (n=5), 16.67\% in Applied Sciences (e.g., Transportation, Meteorology) (n=4), and 12.50\% in Education and Social Sciences (n=3).

An initial pilot study with \textit{3 participants} helped refine our methodology and develop a preliminary codebook. The final study was conducted with the remaining \textit{21 participants}. All participants were compensated with a \$60 gift card. As the authors' institution lacks an Institutional Review Board (IRB), the data collection protocol was reviewed and approved by the institution's Ethics Office. The anonymized participant description and the corresponding answer engines interacted with are described in Appendix \ref{appendix-paricipants}.

\subsubsection{Study Part 1: Pre-Study Questionnaire (5 minutes).} Participants completed a questionnaire (exact questions in Appendix~\ref{appendix-pre-survey}) that asked for (1) high-level demographic information, (2) participants' familiarity with answer engines, abd (3) a list of 3-4 specific technical questions in their area of expertise that they could query in a search engine.
The demographic and familiarity questions are analyzed to understand the recruited participants and their experience with answer engines. The technical questions are used to form the queries used in the next part of the study.

\subsubsection{Study Part 2: Expertise Information Retrieval (40 minutes).} During this part, participants go over one technical question at a time from the list they provided in the pre-study questionnaire and alternate between querying it in an answer engine and then in a traditional search engine. As participants review both engines' results, they were asked to ``think aloud'' \cite{norgaard2006usability, mcdonald2012exploring}, articulating their thoughts and reactions. This approach enables us to capture detailed insights into what works and doesn't with each engine, and compare the results of both engines on a concrete query. Participants are encouraged to interact in-depth with the results (including by clicking on links). Once they are done, they proceed with the next question on their list. Participants typically spent 5-10 minutes per question and were able to go through an average of 6 questions in the 40-minute time frame.

\subsubsection{Study Part 3: Debate Information Retrieval (40 minutes).} This part follows a similar structure to Part 2, but uses opinion-based queries, a common use case for search engines \cite{haider2019invisible,sharma2024generative}. Participants start with answering a series of questions measuring their agreement with various socially and politically charged statements which we collected from the ProCon debate website\footnote{https://www.procon.org/}, on a Likert scale from 1 to 5. Based on their responses, we constructed questions that reflected both supportive and opposing viewpoints. For example, if a participant agreed with the statement ``Zoos should exist'', a supportive query is: ``Why should zoos exist?'' and an opposing query is: ``Why should zoos not exist?''. For each issue where a participant expressed a non-neutral opinion, we prepared either a supportive or opposing query, which the participant ran through both the answer engine and the traditional search engine. We alternated between supportive and opposing queries, allowing us to understand how participants interact with information that aligns or diverges from their viewpoints. Additionally, the study examined whether the answer engine's responses influenced participants' opinions or if they perceived any bias, allowing us to better understand how these systems shape user viewpoints and reveal potential biases.

The study's 21 participants were divided into three groups of 7. Each group interacted with a single answer engine: YouChat, Bing Copilot, or Perplexity AI. These platforms were chosen due to their public accessibility as AI-as-a-Service (AIaaS) systems and their marketing as answer engines\cite{lins2021artificial}. For the traditional search engine, Google Search was selected for all participants. We next go over the findings from the 21 completed study sessions.

\subsection{Pre-Survey Questionnaire Analysis} \label{sec:study_pre_questionnaire}

Regarding participants' familiarity with generative AI (GAI), 37.5\% (9/24) of participants use GAI-based applications daily, 29.1\% (7/24) weekly, 25\% (6/24) monthly, and 8.3\% (2/24) a few times per month, confirming that most participants interact frequently with GAI.

Regarding the use of answer engines specifically, 70.83\% (17/24) of participants were familiar with these systems, 41.16\% use them multiple times per week, and 58.84\% at least once a month. Participants utilized answer engines to conduct research, brainstorm, plan, learn new skills, and obtain faster results compared to traditional search engines. For example, P4 noted, ``\textit{My experience with current answer engines is similar to using a traditional one such as Google. I think it's more handy,}'' while P20 shared, ``\textit{I use it to get improved, accurate, and clear answers to questions, especially regarding my research studies,}'' highlighting their relevance in professional and personal contexts.

\subsection{Sociotechnical Audit of Answer Engine Shortfalls} \label{sec:study_qualitative}

\begin{table*}[ht]
    \resizebox{\textwidth}{!}{
    \begin{tabular}{p{5cm}p{5cm}p{5cm}p{5cm}}
    \toprule
    \textbf{\textcolor{colorans}{\symbolimg{Images/icons/answer_text_color} Answer Text}} & \textbf{\textcolor{colorcit}{\symbolimg{Images/icons/citation_color} Citation}}  & \textbf{\textcolor{colorsrc}{\symbolimg{Images/icons/sources_color} Sources}}  & \textbf{\textcolor{colorint}{\symbolimg{Images/icons/interface_color} User Interface}} \\ \midrule
    \textbf{A.I} Need for objective details in generated answers (21/21)  & \textbf{C.I} Misattribution and misinterpretation of sources cited (21/21)   & \textbf{S.I} Low Frequency of Sources Used for Summarization (19/21) &\textbf{U.I } The lack of selection, and filtering of sources (17/21) \\\midrule
    \textbf{A.II} Lack of holistic viewpoints for opinionated or charged questions (19/21)   & \textbf{C.II} Cherrypicking information based on assumed context (19/21)  & \textbf{S.II} More sources retrieved than used for generating the actual answer (13/21)  &\textbf{U.II}   Lack of human input in generation and source selection (17/21)    \\\midrule
    \textbf{A.III} Overtly confident language while presenting claims (16/21) & \textbf{C.III} Missing citations for claims and information generated (18/21)  & \textbf{S.III} Lack of trust in sources used by the answer engine (12/20) &\textbf{U.III} Answer engines take additional work to verify and trust (14/21) \\\midrule
    \textbf{A.IV} Simplistic language and a lack of creativity and critical thinking (14/21) &  \textbf{C.IV} Transparency of source selection in model responses (15/21)  & \textbf{S.IV} Redundancy in source citation and duplicate content retrieved (12/21)  &\textbf{U.IV} Citations formats are not a normalized interaction (12/21)    \\
    \bottomrule                                
    \end{tabular}}
    \caption{\textcolor{black}{Summary of key findings, organized thematically by answer engine components, with the number of participants who explicitly identified and expressed concerns for each component.}}
    \label{tab:findingsquickSummary}
\end{table*}

We employed a constructivist grounded theory using a qualitative coding approach \cite{charmaz2017constructivist} to analyze our user interview data. Following \citet{charmaz2006constructing}, the authors individually coded the transcripts line-by-line, employing inductive reasoning to develop theories \cite{glaser1967discovery,glaser1992basics}. Using the qualitative coding platform Taguette\footnote{\url{https://app.taguette.org/}}, we generated themes from the transcript snippets. These themes were then synthesized and refined through collaborative discussions between the authors, and insights were categorized with respect to the four components of an answer engine: (1) \textcolor{colorans}{\textbf{answer text}}, (2) \textcolor{colorcit}{\textbf{citation}}, (3) \textcolor{colorsrc}{\textbf{sources}}, and (4) \textcolor{colorint}{\textbf{user interface}}. The final 16 findings are summarized in Table~\ref{tab:findingsquickSummary} which the next sections go over in detail.


\subsubsection{\textbf{\textcolor{colorans}{\symbolimg{Images/icons/answer_text_color} Theme 1: Answer Text Findings}}} 

\textbf{\textcolor{colorans}{A.I. Need for Objective Detail in Generated Answers \textit{(21/21)}:}} A pervasive issue across all three answer engines was the lack of detail and contextual depth in generated responses. This shortfall affected both expertise and debate queries. Participants repeatedly found the summaries to be overly generic and insufficient, often driving them to seek more comprehensive information through traditional search engines like Google.

Participant P1 noted, \textit{``It was just trying to answer without actually giving me a solid answer or a more thought-out answer, which I am able to get with multiple Google searches."} P10 emphasized, ``It's too short and just summarizes everything a lot. [The model] needs to give me more data for the claim, but it's very summarized.'' These reflections highlight a common issue: a desire for answer conciseness leads to \textit{frequently omitted critical details that would substantiate claims}. As a result, responses are perceived as superficial, \textbf{lacking the necessary specificity and nuance}. Participants also expressed concerns about the absence of `numbers' or `percentages' where and when required. As P2 explained, \textit{``There's no quantitative numbers, and I would like to see that if it's citing sources. It should provide metrics maybe when it's making these declarations.''} 



\begin{figure*}[h]
  \centering
  \includegraphics[scale = 0.24]{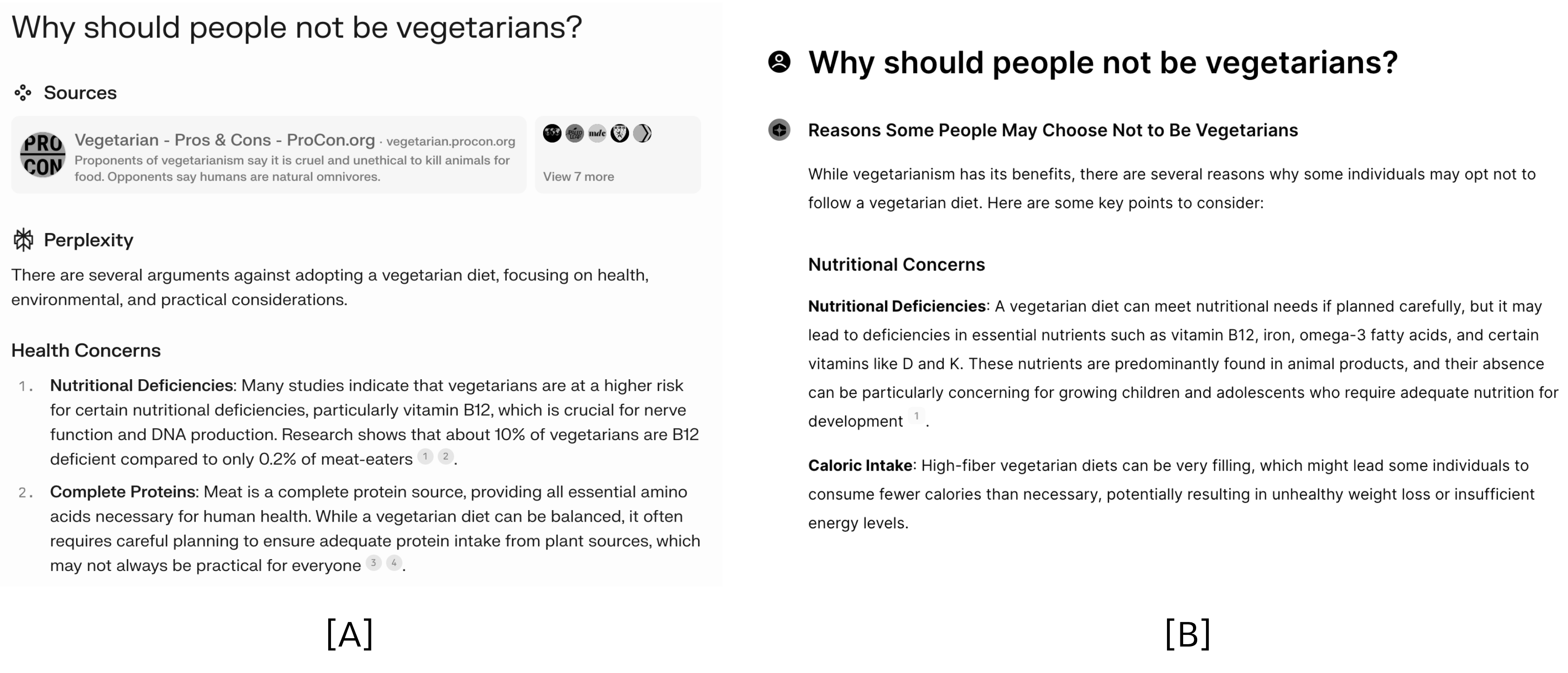}
  \caption{Comparison of outputs from \textbf{[A]} Perplexity, which reflects the bias inherent in the question by presenting only a one-sided response, and \textbf{[B]} YouChat, which acknowledges multiple perspectives, avoiding presenting incomplete information.}
  \label{fig:Opinion}
\end{figure*}

\textbf{\textcolor{colorans}{A.II. Lack of Holistic Viewpoint \textit{(19/21)}:}}
Our study revealed a limitation in the behavior of answer engines when participants engaged with \textit{opinionated queries}. The answer engines frequently aligned with the bias implied in the question, \textbf{neglecting to present diverse perspectives} available from the retrieved sources. The responses often appeared to support only the side of the argument the model inferred the participant was ``looking for,'' thereby \textbf{reinforcing user biases}. Figure~\ref{fig:Opinion} illustrates this by showing an example of a one-sided answer (left, Perplexity.ai) and a comparably more nuanced answer (right, You.com).

Participant P19 noted, ``\textit{I want to find out more about the flip side of the argument... this is all with a pinch of salt because we don't know the other side and the evidence and facts.}'' Similarly, P1 stated, ``\textit{It felt like it was trying to just conform to my question, even though the sources that it was citing actually spoke about all the negatives and positives}," indicating a mismatch between the source content and the generated answer. P4 observed, ``\textit{It is not giving you both sides of the argument; it's not arguing with you. Instead, [the model] is just telling you, 'you're right... and here are the reasons why.'}" These responses highlight a widespread concern: the system's failure to provide a balanced viewpoint, instead mirroring the biases implicit in the questions posed. 



\textbf{\textcolor{colorans}{A.III. Confident Language While Presenting Claims \textit{(16/21)}:}}

Another prominent issue identified by participants is the generation of \textbf{overly confident responses}. All three systems frequently used terms of affirmation and certainty, \textit{even when addressing subjective opinions or claims}. This approach can lead participants to trust the generated content without the necessary context, with the problem being highlighted for both debate and expertise queries. Figure~\ref{fig:Opinion} illustrates the issue: in \textbf{[A]} the answer engine confidently represents information without stating the nuances.

Participants highlighted this issue through their interactions with the models. P3 observed that \textit{``the model uses a very magnetic or authoritative voice while making claims,''} which \textit{``can definitely convince someone that this is `the answer' instead of actually giving them the opportunity to see the issue.''} Similarly, P4 noted that the model employs the `God Voice', likening it to articles that \textit{``make you think that it's the ultimate truth.''} P3 also mentioned, \textit{``It writes so confidently, I feel convinced without even looking at the source. But when you look at the source, it's bad and that makes me question it again.''} P14 provided a social perspective, stating, ``\textit{If someone doesn't exactly know the right answer, they will trust this even when it is wrong.}''




\textbf{\textcolor{colorans}{A.IV. Overly Simplistic Writing Form and a Lack of Critical Thinking \textit{(14/21)}:}}

The fourth finding highlighted by many participants is the simplistic nature of the language used in responses. Participants noted that this simplicity in language reflects a \textbf{lack of critical analysis and thinking}. 

For example, P13 found the answers to be \textit{``kind of childish,''} noting they did not match the scientific level required. P2 described the text as \textit{``fluffy''} and \textit{``similar to what a fifth grader might write without consulting sources''}. P5 quoted \textit{`If I was grading a student's assignment and they had given me that answer... I don't know that I would want to pass a student.''}.

Additionally, some participants perceived the systems as \textbf{‘people pleasers’}, presenting information in a manner that was agreeable or comforting rather than providing a comprehensive or accurate response. P1 noted, \textit{``It’s being a people pleaser and only trying to validate me.''} These insights highlight the need for answer engines to provide more nuanced and detailed responses.

\subsubsection{\textbf{\symbolimg{Images/icons/citation_color} \textcolor{colorcit}{Theme 2: Citation}}}


\textbf{\textcolor{colorcit}{C.I. Misattribution of Sources: Correct Summaries, Incorrect Citations \textit{(21/21)}:}}

A common issue identified in this theme was the \textbf{misattribution of sources}. This occurs when the answer engine cites a source that does not factually support the cited statement, misrepresenting the source content. 
For instance, P12 noted, ``\textit{It has cited irrelevant parts of the paper for this question.}'' P15 commented, ``\textit{But this statement doesn't seem to be in the source. I mean the statement is true; it's valid... but I don't know where it's even getting this information from.}''
Similarly, P17 observed, ``\textit{So [the answer] mentions information on nutrients, but that is not present at all in the whole sentence and the whole article [cited].}''

Participants felt that the systems were \textbf{using citations to legitimize their answer}, creating an illusion of credibility. This facade was only revealed to a few users who proceeded to scrutinize the sources. P4 expressed concern, ``\textit{But does it feel that it's using citations to legitimize itself for its own generation?}” and elaborated, ``\textit{I mean it's just like you see a citation, you assume it's a valid source...I'll just see that there is a source. That's it. I don’t verify it.}” This misattribution highlights a major flaw in the model's process: \textbf{even when a statement is accurate, the citation can point to an irrelevant source}. 



\textbf{\textcolor{colorcit}{C.II. Cherrypicking Information Based on Assumed Context \textit{(19/21)}:}}


When participants posed expert or opinion-based questions, they noticed that the system often selectively presented information from retrieved sources, highlighting \textbf{a particular perspective instead of a comprehensive view of the article}. For instance, P7 noted, ``\textit{I feel [the system] is manipulative. It takes only some information and it feels I am manipulated to only see one side of things.}" Similarly, P8 observed, ``\textit{[The source] actually has both pros and cons, and it's chosen to pick just the sort of required arguments from this link without the whole picture.}" 
P12 added these concerns, stating, "\textit{It's sort of presenting some information that it's picked out already from all the articles and gives that to me. I feel like it sort of reinforces only certain notions.}"

Participants felt that the model does not accurately represent the full scope of the source material. This behavior underscores the model's limitation in potentially misrepresenting sources by taking statements out of context to fit the query. The tendency to reinforce assumed user biases further contributes to an \textbf{echo chamber effect}, limiting exposure to a broader range of perspectives and potentially distorting the original intent of the source material.

\textbf{\textcolor{colorcit}{C.III. Missing Citation for Claims and Information Generated \textit{(18/21)}:}}

\begin{figure*}[h]
  \centering
  \includegraphics[scale = 0.23]{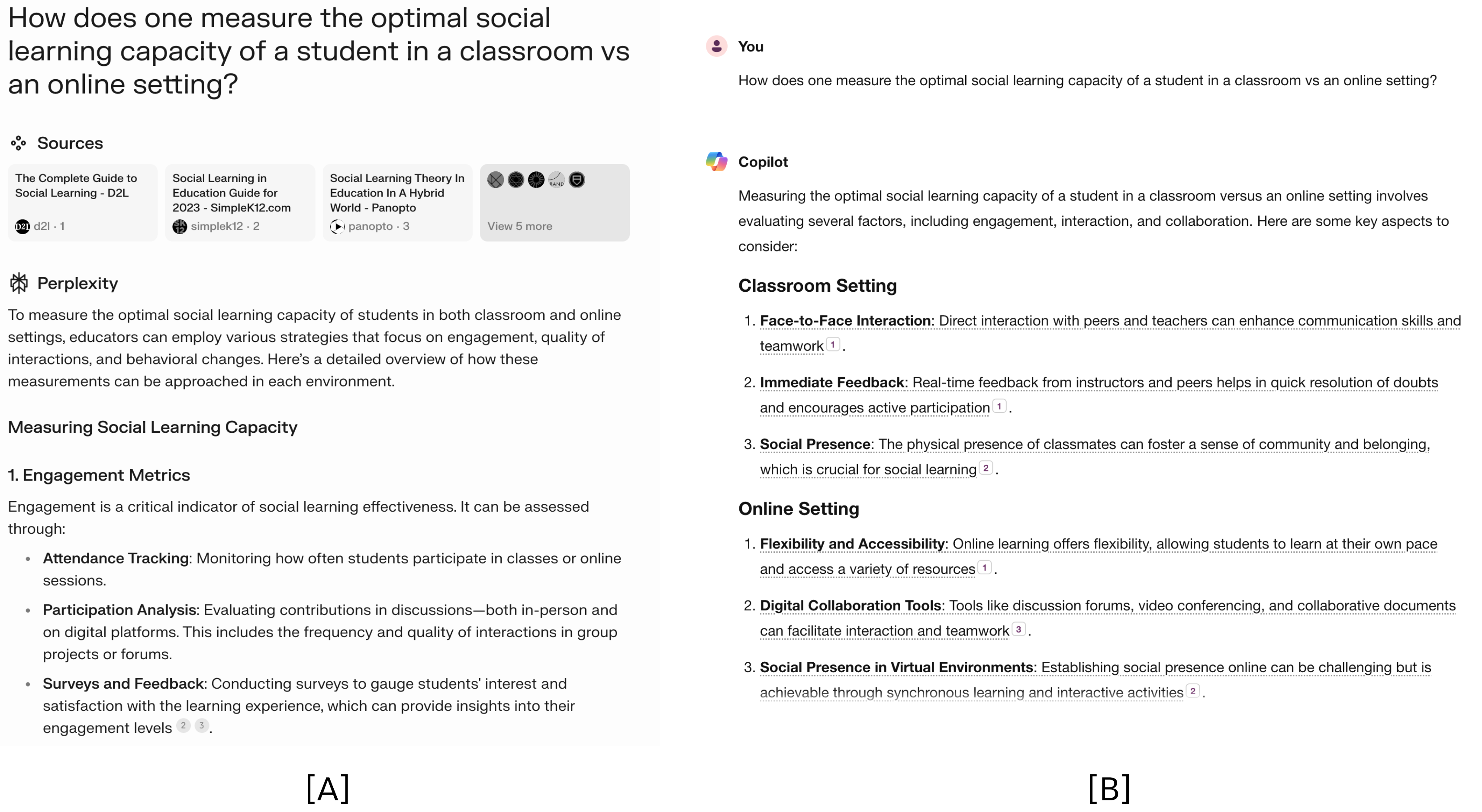}
  \caption{Comparison of outputs from \textbf{[A]} Perplexity, which lacks citations for the points generated, causing confusion on the actual source of each sentence, and \textbf{[B]} Copilot, which effectively indicates the sources for each statement.}
  \label{fig:Citations}
\end{figure*}

The absence of citations in many of the sentences generated by \textit{all three answer engines} emerged as a critical issue, especially when \textbf{key claims or facts are presented without necessary factual support from the sources}. 
P8 expressed frustration with this issue, stating, ``\textit{I really wanted a reference with this [claim] because it is like giving a statement mentioning that this is historical but this is without any citations or without any validation.}" Similarly, P16 highlighted the inconvenience caused, noting, ``\textit{Not having the references for each sentence is annoying...you want to know what's the resource retrieved is. Now we'll have to actually go to the website and compare notes, which is an additional step which no one wants to do. I would have gone to the website in the first place instead.}" 
These comments reveal a clear demand for citations, particularly for sentences that are critical to answering the question. Figure~\ref{fig:Citations} shows an example of an uncited statement in a Perplexity answer, compared to a similar Copilot cited answer.

\textbf{\textcolor{colorcit}{C.IV. Lack of Transparency of Source Selection in Model Responses \textit{(15/21)}:}}

Participants raised significant concerns about the \textbf{lack of transparency in how the system selects and prioritizes citation}, highlighting the need for a clearer explanation of its decision-making process. This ``\textit{black box}" nature of the system makes it difficult for users to trust the outputs, as they cannot discern why certain sources are cited over others.
Participants frequently noted that the system did not adequately prioritize important or factual sources, leading to a general de-emphasis of critical references. 
For instance, P4 questioned, ``\textit{Where is it getting this thing from? Why is it getting it from these particular sources is what I'm curious about.}" 

Additionally, P2 expressed frustration with the system's opaque process on the lack of stating which specific part the information came from the source, stating, ``\textit{It’s very easy to just cough out sources and be like, ‘this is where I took all this information from.’ But which part of the information did you take from? That kind of explanation doesn't exist.}"

Such feedback underscores the importance of transparency in the system's source evaluation criteria. As P5 remarked, ``\textit{It's picking results from the first page of Google Search... but I didn't see any consistency,}" suggesting that \textbf{users need more clarity to trust the system's outputs fully}.

\subsubsection{\textbf{\symbolimg{Images/icons/sources_color} \textcolor{colorsrc}{Theme 3: Sources}}}

\textbf{\textcolor{colorsrc}{S.I. Low Number of Sources \textit{(19/21)}:}}

For both expert and opinionated questions, participants encountered experiences where they \textbf{needed more sources to address the question} at hand. Participants highlighted this issue with specific feedback. For instance, P16 remarked, ``\textit{It feels like it's pulling all of this from one source},'' while P1 noted, ``\textit{Again, everything is extracted from the same source, which is very weird.}'' This indicated a pattern where the answer engine heavily relied on a limited number of sources, \textit{averaging to three sources used}, often leading to incomplete answers.

Interestingly, this issue also caused a phenomenon where the model \textbf{overtly emphasized certain sources for generation}. Participants flagged this as a consequence of the models using very few sources for their answer. P5 mentioned, ``\textit{If it's like citing the right review paper, it can get away with citing only one [source], but it isn't doing that and citing one weak article throughout,}'' and P11 added, ``\textit{It feels off as it's just paraphrasing that one source.}'' By relying on a narrow selection of sources, the answer engines fail to capture the full spectrum of information necessary for a well-rounded answer.

\textbf{\textcolor{colorsrc}{S.II. More Sources Listed Than Used (13/21):}}

Participants using \textit{Bing Copilot and Perplexity} noted that these systems often \textbf{listed multiple retrieved sources without actually citing them in the generated answer}, a behavior described as ``buffing''—creating an impression of thoroughness without substance. This practice led to confusion and eroded trust, as users saw citations that were not integrated into the generated response. For example, P12 remarked, ``\textit{It's giving the impression of multiple sources, but it's just citing something that has a blurb citing to the other source. So it's really just coming off of this one source.}” Similarly, P20 observed: ``\textit{Even when it lists other sources that might be relevant or have other viewpoints, it doesn't actually say much about them.}”

This selective use of sources caused frustration, as participants believed the models ignored more reliable or relevant articles, diminishing the perceived quality of the generated content. P5 expressed this concern, stating, ``\textit{Why didn't they use this [listed but uncited article]? This would be much more reliable relatively!}” Notably, however, YouChat did not display this specific weakness and consistently cited all listed sources in its responses.

\textbf{\textcolor{colorsrc}{S.III. Lack of Trust in Source Types (12/20):}}

The answer engines retrieve content from varying sources, including forums, blogs, opinion pieces, and research articles. However, participants expressed general distrust toward certain sources due to perceived biases, or lack of authority. This distrust was evident in feedback like P1's remark, ``\textit{Who knows who has written that post [...] it's a LinkedIn post when the question is a scientific one,}" and P2's observation, ``\textit{The sources are not convincing. They seem to be these non-factual sources from where this answer is kind of drawn.}" 
Additionally, participants noted issues with \textbf{outdated or misinformed content} being used to generate answers. For example, P10 mentioned, "\textit{I think the citation is outdated... because it says 'Windows 10', and we've already switched to better OS, which is what the answer needed}".

\textbf{\textcolor{colorsrc}{S.IV. Different Sources but Duplicate Content (12/21):}}

Participants identified instances where the answer engine retrieved and cited multiple sources that, upon closer inspection, contained identical or highly similar content. While these sources were presented as distinct entities with different citation numbers, they \textbf{ultimately contributed redundant information}.
Such findings raised concerns about potential inaccuracies in the system's source selection process. For example, P3 observed, ``\textit{Source 1 and 2 are just the same content in different formats. This is funny as it’s using them differently,}''
and P9 stated, ``\textit{I think this is a big problem! I should have checked the citations on the other answer as well because it’s basically just giving me one website but citing it differently multiple times.}''

\begin{figure*}[h]
  \centering
  \includegraphics[scale = 0.23]{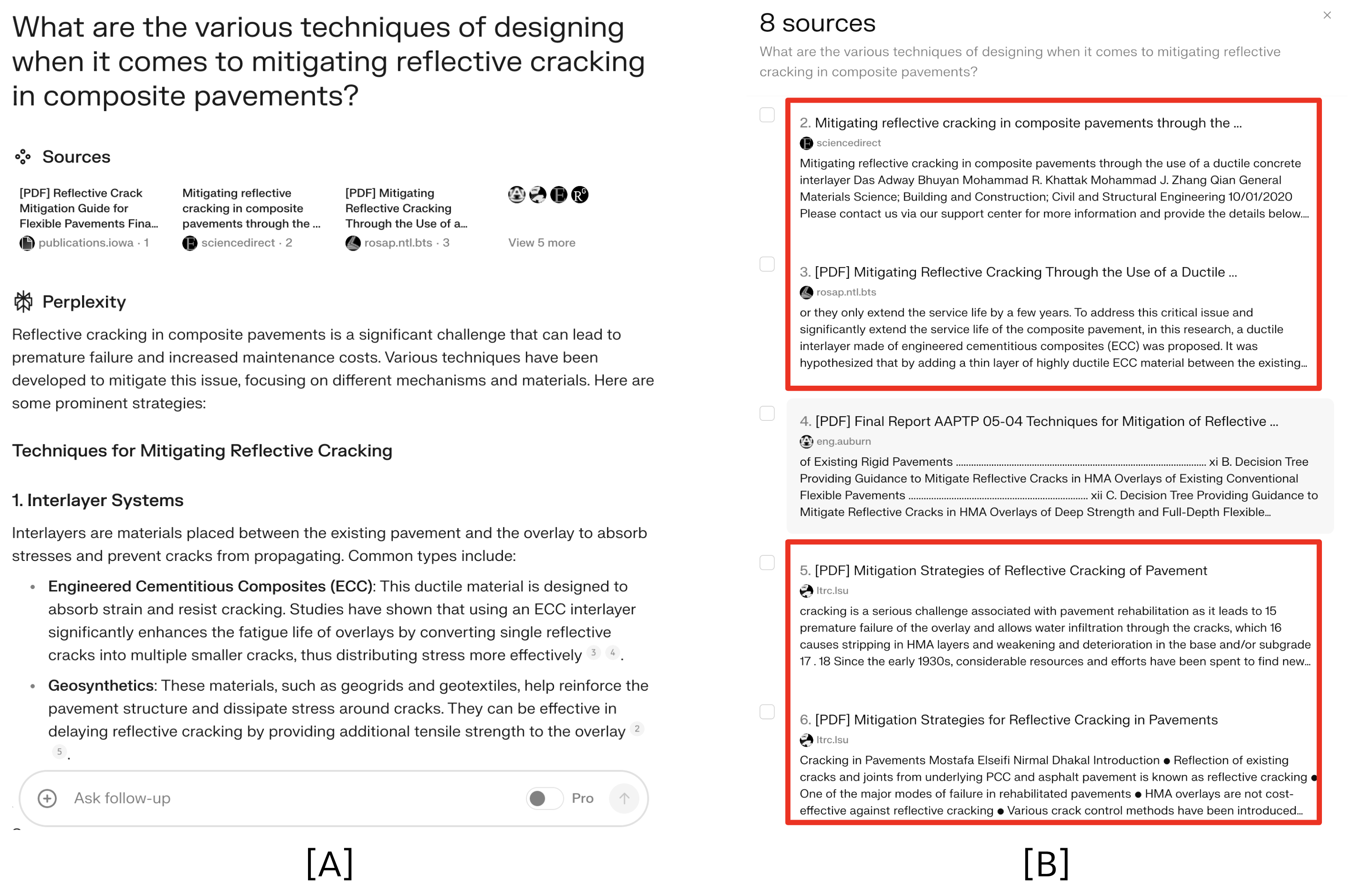}
  \caption{Results generated by Perplexity \textbf{[A]} and the corresponding sources retrieved \textbf{[B]}. The image illustrates how the model retrieved 8 sources, many of which are duplicates of the same source. Despite this, the model cites them differently, creating an illusion of varied content when it is actually the same.}
  \label{fig:Duplicate}
\end{figure*}

This phenomenon reveals limitations in the model's sourcing strategy, suggesting a reliance on superficial differences, such as formatting or minor variations, to justify multiple citations. As a result, the system provides a misleading appearance of a well-rounded answer while simply recycling the same content. Figure~\ref{fig:Duplicate} provides an example of duplicate sourcing.

\subsubsection{\textbf{\textcolor{colorint}{\symbolimg{Images/icons/interface_color} Theme 4: User Interface}}}

\textbf{\textcolor{colorint}{U.I. Autonomy over Source Validation and Selection \textit{(17/21)}:}}

Participant feedback underscored that answer engines often offer narrow perspectives, due to users having little to no control over the information presented to them, leading to concerns about a lack of autonomy and the inability to evaluate the credibility of sources independently. P5 stated, ``\textit{The [answer engine] did not give me a whole lot of confidence in its ability to assess the quality of sources. I would have more trust in my ability to do that.}'' This sentiment suggests that users feel more competent in selecting and scrutinizing sources themselves. 
P11 further elaborated, “\textit{[The answer engine] is taking from these sources, but are they even legit? I cannot choose or remove the ones it chose. At least with Google, I have a lot more choices to click.}”

These findings suggest that while answer engines may provide quick responses, they fail to meet user expectations for transparency and control. The inability to select, filter, or assess sources independently not only limits the breadth of information but also undermines user confidence in the accuracy and reliability of the answers provided.

\textbf{\textcolor{colorint}{U.II. Lack of Human Input in Generation and Source Selection \textit{(17/21)}:}}

Our study identified a key issue with answer engines: the \textbf{tendency to lose context when generating responses} due to the answer engine assuming the most probable context rather than accurately interpreting or clarifying the specific context. 
Participant P7 highlighted this issue: ``\textit{I think the sources were right, but the context is lost. It did not manage to go to the expected answers at all.}” This points to a core limitation: even with accurate sources, the inability to maintain the correct context of the query results in unsatisfactory answers. P9 further noted, “\textit{This is a very generic answer that just has basic features [of the answer]. It doesn't say anything at all of the context asked.}”
Participants suggested that the single-interaction design contributes to this issue. In cases where context is critical, participants recommended that \textbf{the system should adopt a more interactive, conversational approach by asking questions before answering}. By asking clarification questions, answer engines could better address ambiguity and provide more accurate responses. P1 also remarked, "\textit{Having [the system] ask further questions to alleviate any potential ambiguity on the generation would have helped, instead of assuming the context.}"


\textbf{\textcolor{colorint}{U.III. Additional Work to Verify and Trust Answers \textit{(14/21)}:}}

As described in previous findings, participants often felt compelled to independently verify the provided sources due to distrust. 
Therefore, contrary to their intended purpose, some participants found that \textbf{the answer engines often resulted in more work for users, undermining their marketed goal of simplifying information retrieval}.

Participant P2 highlighted this inefficiency: ``\textit{Even if it's gonna give me five paragraphs with 300 sources, it's not going to help me because [...] I'm gonna go and check each source, so this is just another level of information interaction for me.}” The necessity to cross-check each source individually added complexity, extending the time required to complete tasks that might have been faster with a traditional search engine. P17 echoed this, stating, ``\textit{We have to go to the website and compare notes and all of that [for verification], which is an additional step which no one wants to do. I would have just gone to the website in the first place.}” P5 also emphasized the inefficiency: ``\textit{The [answer engine] does not speed up the process. I don't feel like I would trust it enough to just go off that.}” 

\textbf{\textcolor{colorint}{U.IV. Citation Formats are Not a Normalized Interaction \textit{(12/21)}:}}
The format of citations affects user experience. The most common method involves numerical citations (e.g., [1], [1][2]), where numbers correspond to references listed below the content. While familiar to those accustomed to academic writing, this format can be \textbf{less intuitive or effective for individuals who do not regularly engage with such reference systems}. Participant P10 highlighted this issue: ``\textit{I actually really don’t understand the citations because in my job and daily life, it isn’t used.}” This concern suggests that numerical citations, while clear to some, may not be meaningful to a general audience unfamiliar with academic or research practices. P14 echoed this sentiment, noting, ``\textit{People may not understand how [citations] work. People rarely, say my mom or aunt, will not understand what to make of it or how that works as well.}”

However, participants who interacted with Bing Copilot, which uses a hover-over feature to display source information, reported a better understanding of where the information came from. This feature allows users to hover over sentences to see the corresponding source, \textbf{encouraging direct interaction with the sources in a more accessible manner}. Participant P2 commented, ``\textit{[The on-hover] is very useful to have. Sources here, I think it will help more people to feasibly check for sources every time they're taking something from a system like this as they are forced to interact with it.}” This suggests that simple numerical references may not be sufficient to build trust and that more interactive approaches could enhance user engagement and comprehension.

We have identified the major findings from our participant interactions in this section. Additionally, minor findings are included in Appendix \ref{appendix-additional-findings}.

\subsection{Source and Citation Interaction Evaluation} \label{sec:study_quantitative}


To complement qualitative insights, we investigate the number of sources displayed and cited by the three answer engines and participants' interaction with these sources. Our study revealed that the chosen answer engines cite a limited subset of sources displayed. Table \ref{table:citations_used} illustrates the disparity between sources retrieved \textbf{(mean: 4.31)} and sources cited \textbf{(mean: 3.0)} in the final answers, varying across answer engines. The analysis quantitatively supports the finding that answer engines display more sources than they cited, causing user confusion and distrust.

Out of the three answer engines, Perplexity displays the most sources (mean: 5.00), but cites the least (mean: 2.58), while YouChat cites all the sources that are displayed (mean: 3.57). With \textit{an average of only three sources available}, users have limited autonomy in selecting and verifying information.

\begin{table*}[]
\small
\begin{tabular}{ccccclcccc}
\hline
\textbf{} & \multicolumn{4}{c}{\textbf{Sources Displayed by Answer Engine}} &  & \multicolumn{4}{c}{\textbf{Sources Cited by Answer Engine}} \\ \cline{1-5} \cline{7-10} 
\textbf{} & \textbf{All} & \textbf{Perplexity} & \textbf{Bing Copilot} & \textbf{YouChat} &  & \textbf{All} & \textbf{Perplexity} & \textbf{Bing Copilot} & \textbf{YouChat} \\ \cline{1-5} \cline{7-10} 
\textbf{Mean} & \textbf{4.31} & \textbf{5.00} & \textbf{4.46} & \textbf{3.57} &  & \textbf{3.00 (69\%)} & \textbf{2.58 (51\%)} & \textbf{2.80 (62\%)} & \textbf{3.57 (100\%)} \\
\textbf{Median} & 5.00 & 5.00 & 4.00 & 4.00 &  & 3.00 & 3.00 & 3.00 & 4.00 \\
\textbf{SD} & 1.32 & 0.00 & 1.56 & 1.25 &  & 1.15 & 0.65 & 1.18 & 1.25 \\
\textbf{Max} & 8.00 & 5.00 & 8.00 & 6.00 &  & 6.00 & 4.00 & 6.00 & 6.00 \\
\textbf{Min} & 1.00 & 5.00 & 1.00 & 1.00 &  & 1.00 & 1.00 & 1.00 & 1.00 \\ \hline
\end{tabular}
\caption{Sources Retrieved vs Cited by the three answer engines evaluated in the study. The percentage of sources cited is mentioned in `( )' to identify the subset of sources actually cited or used for the generated answer, for each answer engine.}
\label{table:citations_used}
\end{table*}



\begin{figure*}[h]
  \centering
  \includegraphics[scale = 0.20]{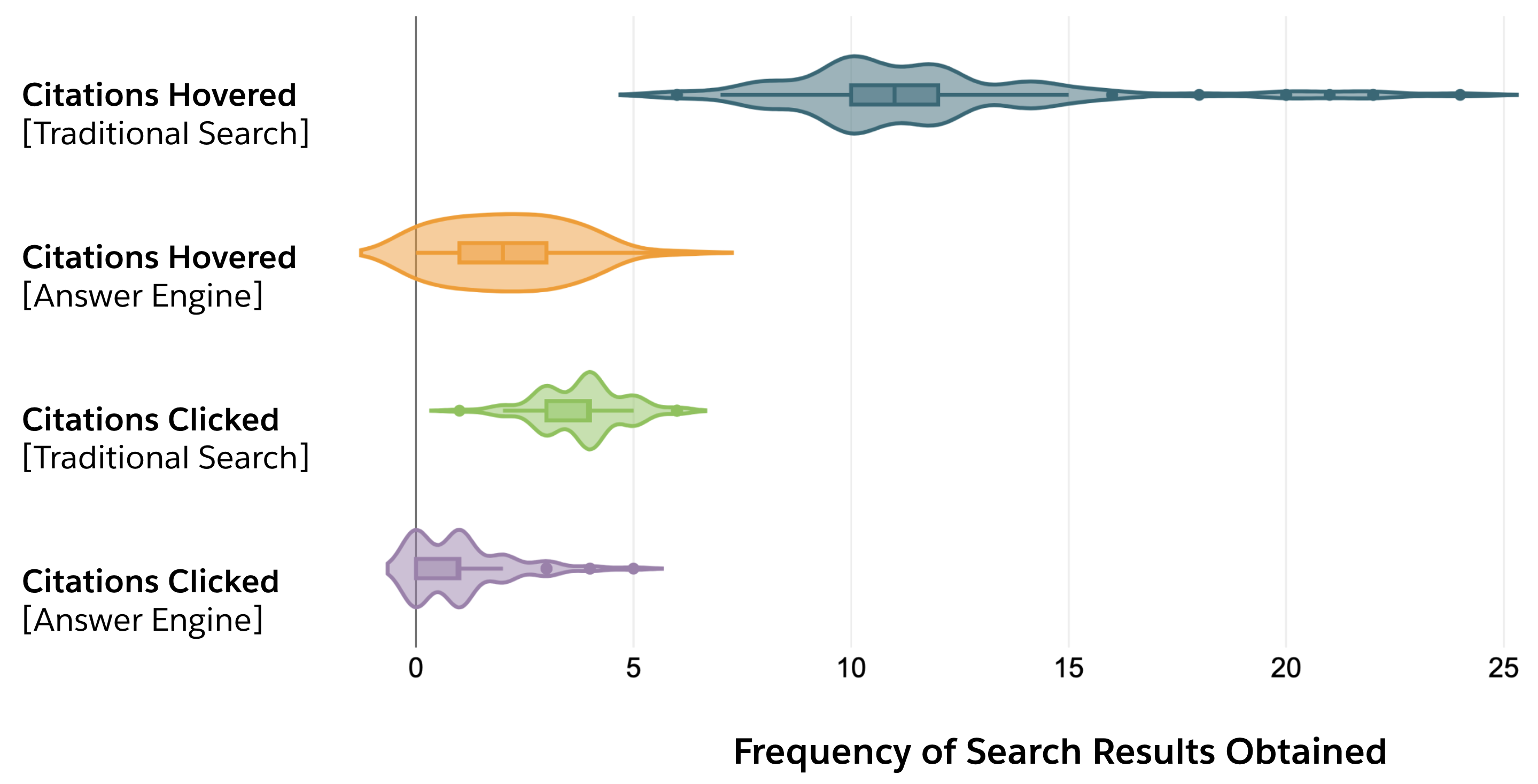}
  \caption{Violin plot showcasing the distribution of number of sources hovered and clicked on by participants for Traditional Search versus Answer Engines.}
  \label{fig:Violin}
\end{figure*}

Further investigation into user interaction with sources reveals a stark contrast in how individuals hover and click sources when using answer engines versus traditional search engines. As shown in Table \ref{table:citations_parsed} and Figure \ref{fig:Violin}, participants display a much narrower scope of inquiry when leveraging answer engines, hovering an average of only 2 sources \textit{(median: 2; SD: 1.39)}. This limited engagement likely results from the fewer sources provided by answer engines, which restricts the breadth of information users can explore. In contrast, participants using traditional search engines adopt a more comprehensive approach, hovering an average of 12 sources \textit{(median: 11; SD: 3.56)}. Additionally, there is a notable difference in the number of sources clicked to thoroughly analyze to find answers. With answer engines, users tend to click on a single source and trust the model's selection. However, when using traditional search engines, participants engage with a wider range of sources, clicking and analyzing an average of 4 sources.

We conducted independent two-sample t-tests to statistically assess our findings, comparing user interactions during traditional search engine use with those observed when using each answer engine. The significance level was set to $\alpha$ = 0.01. As shown in Table \ref{table:ttest}, the results reveal statistically significant differences in user interactions, allowing us to reject the null hypothesis that answer engines foster similar interactions as traditional engines. 


\begin{table*}[]
    \small
    \begin{tabular}{ccccclcccc}
    \hline
     & \multicolumn{4}{c}{\textbf{Citations Hovered by Users}} &  & \multicolumn{4}{c}{\textbf{Citations Clicked by User}} \\ \cline{1-5} \cline{7-10} 
     & \textbf{Perplexity} & \textbf{Bing Copilot} & \textbf{YouChat} & {\color[HTML]{009901} \textbf{Google}} &  & \textbf{Perplexity} & \textbf{Bing Copilot} & \textbf{YouChat} & {\color[HTML]{009901} \textbf{Google}} \\ \cline{1-5} \cline{7-10} 
    \textbf{Mean} & 2.12 & 2.10 & 2.00 & {\color[HTML]{009901} \textbf{11.81}} &  & 1.29 & 0.76 & 1.07 & {\color[HTML]{009901} \textbf{3.80}} \\
    \textbf{Median} & 2.00 & 2.00 & 2.00 & {\color[HTML]{009901} \textbf{11.00}} &  & 1.00 & 1.00 & 0.50 & {\color[HTML]{009901} \textbf{4.00}} \\
    \textbf{SD} & 1.39 & 1.21 & 1.72 & {\color[HTML]{009901} \textbf{3.56}} &  & 0.90 & 0.89 & 1.41 & {\color[HTML]{009901} \textbf{0.96}} \\
    \textbf{Max} & 5.00 & 4.00 & 6.00 & {\color[HTML]{009901} \textbf{24.00}} &  & 3.00 & 4.00 & 5.00 & {\color[HTML]{009901} \textbf{6.00}} \\
    \textbf{Min} & 0.00 & 0.00 & 0.00 & {\color[HTML]{009901} \textbf{6.00}} &  & 0.00 & 0.00 & 0.00 & {\color[HTML]{009901} \textbf{1.00}} \\ \hline
    \end{tabular}
    \caption{Citations Hovered versus Clicked by a participant in a traditional search engine and the chosen answer engines. The tradiditional search engine results are \color[HTML]{009901} \textbf{highlighted} to differentiate the system and its result.}
    \label{table:citations_parsed}
\end{table*}

\begin{table*}[]
\small
\begin{tabular}{ccccclcccc}
\hline
 & \multicolumn{4}{c}{\textbf{Citation Hovered by Users}} &  & \multicolumn{4}{c}{\textbf{Citations Clicked by User}} \\ \cline{1-5} \cline{7-10} 
 & \textbf{T vs All} & \textbf{T vs Perp} & \multicolumn{1}{l}{\textbf{T vs Copilot}} & \multicolumn{1}{l}{\textbf{T vs You}} &  & \textbf{T vs All} & \textbf{T vs Perp} & \multicolumn{1}{l}{\textbf{T vs Copilot}} & \multicolumn{1}{l}{\textbf{T vs You}} \\ \cline{1-5} \cline{7-10} 
\textbf{t-statistic} & 22.94 & 13.00 & 14.59 & 14.00 & \multicolumn{1}{c}{} & 17.13 & 11.40 & 15.06 & 11.43 \\
\textbf{pvalue} & 8.14e-53 & 1.56e-23 & 1.73e-27 & 5.06e-26 & \multicolumn{1}{c}{} & 3.21e-28 & 5.08e-20 & 1.66e-28 & 2.50e-20 \\ \hline
\end{tabular}
\caption{Independant two sample t-test between the citations hovered and clicked between traditional search engine \textbf{(T)} and each of the answer engines (All Answer Engine \textbf{(All)}, Perplexity \textbf{(Perp)}, Bing Copilot \textbf{(Copilot)} and You Chat \textbf{(You)})}
\label{table:ttest}
\end{table*}

Finally, our analysis in Table \ref{table:aligned_disaligned} examined how individuals interact with answer engines when asking questions that align with or challenge their acknowledged stance. 
When faced with contradictory questions, participants engaged in more analysis, hovering (mean = 1.72) and clicking (mean = 2.95) a higher number of sources. In contrast, when asking aligned questions, participants relied on fewer sources (hovering: mean = 1.08; clicking: mean = 0.48) and often trusted information with lesser verification. The two sample t-tests confirmed a significant difference in verification behaviors between the two conditions (hovering; p-value = 4.93e-06, clicking; p-value = 4.88e-05). 

\begin{table*}[]
\begin{tabular}{ccclcc}
 & \multicolumn{2}{c}{\textbf{Citation Hovered}} &  & \multicolumn{2}{c}{\textbf{Citations Clicked}} \\ \cline{2-3} \cline{5-6} 
 & \textbf{Aligned Q} & \textbf{Disaligned Q} &  & \textbf{Aligned Q} & \textbf{Disaligned Q} \\ \cline{1-1} \cline{2-3} \cline{5-6} 
\textbf{Mean} & 1.08 & 2.95 &  & 0.48 & 1.72 \\
\textbf{Median} & 1.00 & 3.00 &  & 0.00 & 2.00 \\
\textbf{SD} & 1.25 & 1.21 &  & 0.77 & 1.12 \\
\textbf{Max} & 4.00 & 6.00 &  & 3.00 & 5.00 \\
\textbf{Min} & 0.00 & 0.00 &  & 0.00 & 0.00 \\ \hline
\end{tabular}
\caption{Citations hovered versus clicked by a participant in settings where they asked question aligned/disaligned from their acknowledged biases, across all three systems combined.}
\label{table:aligned_disaligned}
\end{table*}

\section{Design Recommendations and Metrics} \label{sec:recs_and_metrics}
\subsection{Answer Engine Design Recommendation}

\begin{table*}[]
    \resizebox{0.98\textwidth}{!}{%
    \renewcommand{\arraystretch}{1.4}
    \begin{tabular}{p{4.5cm}p{7.5cm}c}
        \toprule
        \textbf{Design Recommendation} & \textbf{Associated System Weakness} & \textbf{Metric Developed} \\
        \midrule
        Provide balanced answers & Lack of holistic viewpoints for opinionated questions \textcolor{colorans}{[A.II]} & One-Sided Answers \\
        Provide objective detail to claims & Overly confident language when presenting claims \textcolor{colorans}{[A.III]} & Overconfident Answers \\
        Minimize fluff information & Simplistic language and a lack of creativity \textcolor{colorans}{[A.IV]} & Relevant Statements \\
        Reflect on answer thoroughness & Need for objective detail in answers \textcolor{colorans}{[A.I]} & -- \\ \hline
        Avoid unsupported citations & Missing citations for claims and information \textcolor{colorcit}{[C.III]} & Unsupported Statement \\
        Double-check for misattributions & Misattribution and misinterpretation of sources cited \textcolor{colorcit}{[C.I]} & Citation Accuracy \\
        Cite all relevant sources for a claim & Transparency of source selected in model response \textcolor{colorcit}{[C.IV]} & Source Necessity \\
        Listed \& Cited sources match & More sources retrieved than used \textcolor{colorsrc}{[S.II]}& Uncited Sources \\ \hline
        Give importance to expert sources & Lack of trust in sources used \textcolor{colorsrc}{[S.III]} & Citation Thoroughness \\
        Present only necessary sources & Redundancy in source citation \textcolor{colorsrc}{[S.IV]} & Source Necessity \\
        Differentiate source \& LLM content & More sources retrieved than used for generation \textcolor{colorsrc}{[S.II]} & \_ \\
        Full represent source type & Low frequency of source used for summarization \textcolor{colorsrc}{[S.I]} & \_ \\  \hline
        Incorporate human feedback & Lack of search, select and filter \textcolor{colorint}{[U.I]} & \_ \\
        Implement interactive citation & Citation formats are not normalized interactions \textcolor{colorint}{[U.IV]} & \_\\
        Implement localized source citation & Additional work to verify and trust sources \textcolor{colorint}{[U.II]} & \_ \\
        No answer when info not found  & Lack of human input in generation and selection \textcolor{colorint}{[U.I]} & \_ \\
        \bottomrule
    \end{tabular}
    }
    \vspace{0.5em}
    \caption{Sixteen design recommendations for answer engines. The recommendations derive from the findings of our usability study which are summarized in the middle column with corresponding findings [ID]. Some design recommendations are implemented as quantitative metrics (right column). Appendix~\ref{appendix-recommendations} defines each recommendation.}
    \label{table:recommendations}
\end{table*}

To make findings from the study concrete and actionable, we distill them into a set of design recommendations for answer engine development. When possible, we then tie a recommendation with a computable metric that can measure relative adherence with the metric. Table~\ref{table:recommendations} provides the mapping between the study findings, the design recommendations, and the metrics. A detailed definition of each recommendation is presented in Appendix~\ref{appendix-recommendations}, and we now define each of the metrics.

\subsection{Answer Engine Metrics}

\begin{figure*}[h]
  \centering
  \includegraphics[width=0.95\textwidth]{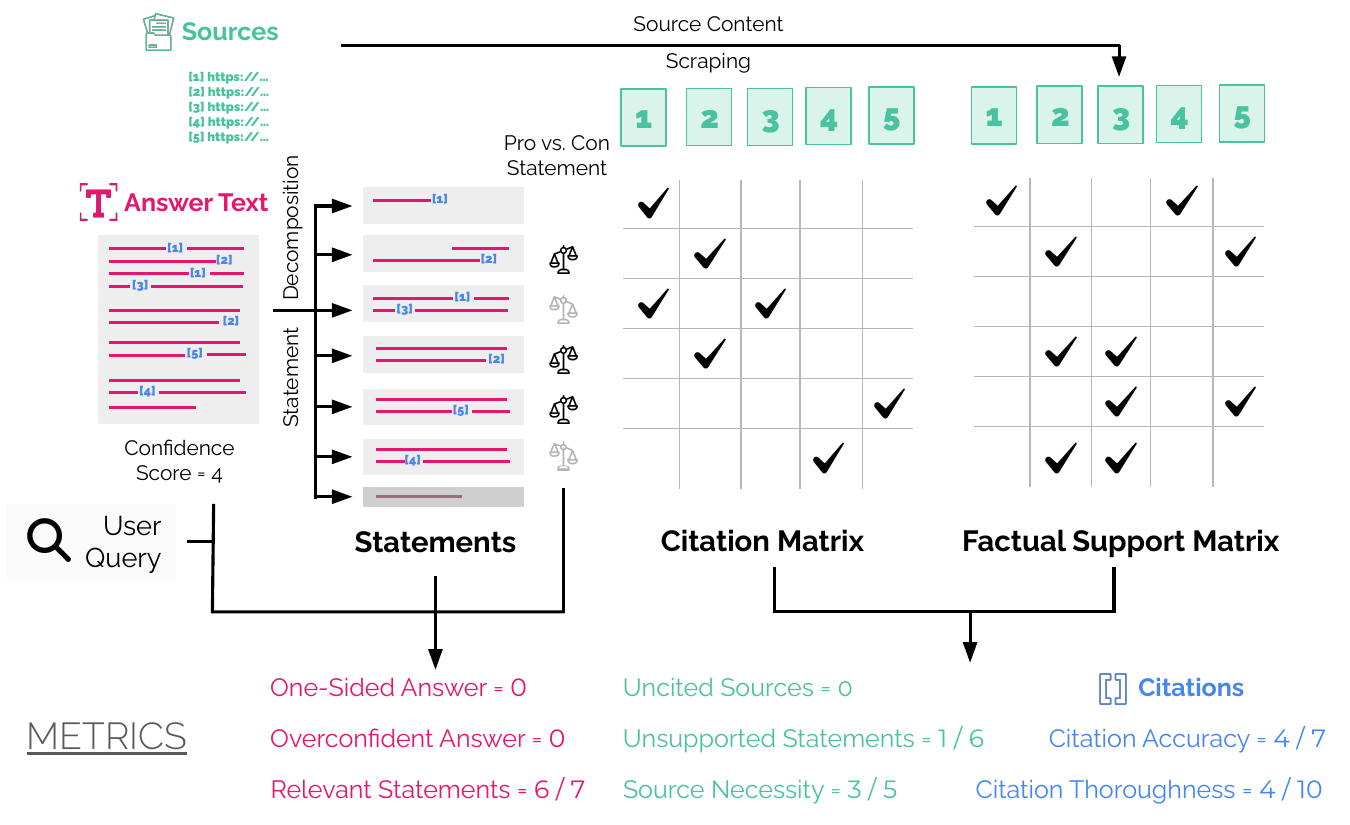}
  \caption{Illustrative diagram of the processing of an answer engine's response into the 8 metrics of the Answer Engine Evaluation Framework. The description of each metrics is illustrated in Section 4.2.}
  \label{fig:scorecard_pipeline}
\end{figure*}

Figure~\ref{fig:scorecard_pipeline} illustrates the processing of an answer engine's response into the 8 metrics of the Answer Engine Evaluation Framework, while Table \ref{table:recommendations} showcases the motivation and rationale behind each metric. We first go over the preliminary processing common to several metrics, then define each metric.

\subsubsection{Preliminary Processing} \label{sec:preliminary_processing}

When evaluating an answer engine, our evaluation framework requires the extraction of four content elements: the user query (1), the generated answer text (2) with the embedded citation (3) to the sources represented by a publicly accessible URL (4). Because APIs made available by answer engines do not provide all of these elements, we implemented automated browser scripts to extract these elements for three popular answer engines: You.com, Perplexity.ai, and BingChat\footnote{Extending the evaluation to other answer engines would require adapting the scripts to the specific website structure of the target answer engines.}. Several operations below rely on LLM-based processing, for which we default to using GPT-4o, and have listed the prompts used in Appendix~\ref{appendix:prompts}. When necessary, we evaluate the accuracy of LLM-based processing and report on the level of agreement with manual annotation.

A first operation consists of decomposing the answer text into statements. A statement is typically a sentence, and occasionally a sub-unit of a sentence. Decomposing the answer into statements allows to study the factual backing of the answer by the sources at a granular level, and is common in fact-checking literature \cite{laban2022summac,tang2024minicheck}. In the example of Figure~\ref{fig:scorecard_pipeline}, the answer text is decomposed into seven statements. Each statement is further assigned two attributes: \textbf{Query Relevance} is a binary attribute that indicates whether the statement contains answer elements relevant to the user query. Irrelevant statements are typically introductory or concluding statements that do not contain factual information (e.g., ``That's a great question!'', ``Let me see what I can do here''). \textbf{Pro vs. Con Statement} is calculated only for biased debate queries and is a ternary label that measures whether the statement is pro, con, or neutral to the bias implied in the query formulation. In the example, the first six statements are relevant, and the last is irrelevant (greyed out), and three statements are pro, two are con, and two are neutral.

A second operation consists of assigning an \textbf{Answer Confidence score} to the answer using a Likert scale (1-5), with 1 representing Strongly not Confident and 5 representing Strongly Confident. Answer confidence is assigned by an LLM judge instructed with a prompt that provides examples of phrases used to express different levels of confidence. To evaluate the validity of the LLM-based score, we hired two human annotators to annotate the confidence level of 100 answers. We observed a Pearson correlation of 0.72 between the LLM judge and human annotators, indicating substantial agreement, and confirming the reliability of the LLM judge for confidence scoring. In the example of Figure~\ref{fig:scorecard_pipeline}, the answer is assigned an illustrative confidence score of 4 (Confident).

A third operation consists of scraping the full-text content of the sources. We leverage Jina.ai's Reader tool\footnote{https://jina.ai/reader/}, to extract the full text of a webpage given its URL. Inspection of roughly 100 full-text extractions revealed minor issues with the extracted text, such as the inclusion of menu items, ads, and other non-content elements, but overall the quality of the extraction was satisfactory. For roughly 15\% of the URLs, the Reader tool returns an error either due to the web page being behind a paywall, or due to the page being unavailable (e.g., a 404 error). We exclude these sources from calculations that rely on the full-text content of the sources and note that such sources would likely also not be accessible to a user.

A fourth operation creates the \textbf{Citation Matrix} by extracting the sources cited in each statement. The matrix (center in Figure~\ref{fig:scorecard_pipeline}) is a (number of statements) x (number of sources) matrix where each cell is a binary value indicating whether the statement cites the source. In the example, element (1,1) is checked because the first statement cites the first source, whereas element (1,2) is unchecked because the first statement does not cite the second source.

A fifth operation creates the \textbf{Factual Support Matrix} by assigning for each (statement, source) pair a binary value indicating whether the source factually supports the statement. We leverage an LLM judge to assign each value in the matrix. A prompt including the extracted source content and the statement is constructed, and the LLM must determine whether the statement is supported or not by the source. Factual support evaluation is an open challenge in NLP \cite{tang2024minicheck,kim2024fables}, but top LLMs (GPT-4o) have been shown to perform well on the task \cite{laban2023llms}. To understand the degree of reliability of LLM-based factual support evaluation in our context, we hired two annotators to perform 100 factual verification tasks manually. We observed a Pearson correlation of 0.62 between the LLM judge and manual labels, indicating moderate agreement. Relying on an LLM to measure factual support is a limiting factor of our evaluation framework, necessary to scale our experiments: we ran on the order of 80,000 factual support evaluations in upcoming experiments, which would have been cost-prohibitive through manual annotation. In the first row of the example Factual Support matrix, columns 1 and 4 are checked, indicating that sources 1 and 4 factually support the first statement.

For the annotation efforts, we hired a total of four annotators who are either professional annotators hired in \textit{User Interviews}\footnote{www.userinterviews.com/}, or graduate students enrolled in a computer science degree. We provided clear guidelines to annotators for the task and had individual Slack conversations where each annotator could discuss the task with the authors of the paper. Annotators were compensated at a rate of \$25 USD per hour.

With the preliminary processing complete, we can now define the 8 metrics of the Answer Engine Evaluation Framework.

\subsubsection{Answer text Metrics}

\textbf{1. One-Sided Answer:} This binary metric is only computed on debate questions, leveraging the Pro vs. Con statement attribute. An answer is considered one-sided if it does not include both pro and con statements on the debate question.

\begin{equation}
    \text{One-Sided Answer} = \begin{cases} 0 & \text{both pro and con} \\ & \text{statements are present} \\ 1 & \text{otherwise} \end{cases}
\end{equation}

In the example of Figure~\ref{fig:scorecard_pipeline}, \texttt{One-Sided Answer = 0} as there are three pro statements and two con statements. When considering a collection of queries, we can compute \texttt{\% One-Sided Answer} as the proportion of queries for which the answer is one-sided.

\textbf{2. Overconfident Answer:} This binary metric leverages the Answer Confidence score, combined with the One-Sided Answer metric and is only computed for debate queries. An answer is considered overconfident if it is both one-sided and has a confidence score of 5 (i.e., Strongly Confident).


\begin{equation}
    \text{Overconfident Answer} = \begin{cases} 
    1 & \text{if One-Sided Answer = 1} \\
      & \text{and Answer Confidence = 5} \\
    0 & \text{otherwise} 
    \end{cases}
\end{equation}

We implement a confidence metric in conjunction with the one-sided metric as it is challenging to determine the acceptable confidence level for any query. However, based on our study findings, an undesired trait in an answer is to be overconfident while not providing a comprehensive and balanced view, which we capture with this metric. In the example of Figure~\ref{fig:scorecard_pipeline}, \texttt{Overconfident Answer = 0} since the answer is not one-sided. When considering a collection of queries, we can compute \texttt{\% Overconfident Answer} as the proportion of queries with overconfident answers.

\textbf{3. Relevant Statement:} This ratio metric measures the fraction of relevant statements in the answer text in relation to the total number of statements.

\begin{equation}
    \text{Relevant Statement} = \frac{\text{Number of Relevant Statements}}{\text{Total Number of Statements}}
\end{equation}

This metric captures the to-the-pointedness of the answer, limiting introductory and concluding statements that do not directly address the user query. In the example of Figure~\ref{fig:scorecard_pipeline}, \texttt{Relevant Statement = 6/7}.

\subsubsection{Sources Metrics}

\textbf{4. Uncited Sources:} This ratio metric measures the fraction of sources that are cited in the answer text in relation to the total number of listed sources.

\begin{equation}
    \text{Uncited Sources} = \frac{\text{Number of Cited Sources}}{\text{Number of Listed Sources}}
\end{equation}

This metric can be computed from the citation matrix: any empty column corresponds to an uncited source. In the example of Figure~\ref{fig:scorecard_pipeline}, since no column of the citation matrix is empty, \texttt{Uncited Sources = 0 / 5}.

\textbf{5. Unsupported Statements:} This ratio metric measures the fraction of relevant statements that are not factually supported by any of the listed sources. Any row of the factual support matrix with no checked cell corresponds to an unsupported statement.

\begin{equation}
    \text{Unsupported Statements} = \frac{\text{No. of Unsupported Statements}}{\text{No. of Relevant Statements}}
\end{equation}

In the example of Figure~\ref{fig:scorecard_pipeline}, the third row of the factual support matrix is the only entirely unchecked row, indicating that the third statement is unsupported. Therefore, \texttt{Unsupported Statements = 1 / 6}.

\textbf{6. Source Necessity:} This ratio metric measures the fraction of sources that are necessary to factually support all relevant statements in the answer text. Understanding what source is necessary or redundant can be formulated as a graph problem. We transform the factual support matrix into a (statement,source) bi-partite graph. Finding which source is necessary is equivalent to determining the minimum vertex cover for source nodes on the bipartite graph. We use the Hopcroft-Karp algorithm \cite{hopcroft1973n} to find the minimum vertex cover, which tells us which sources are necessary to cover factually supported statements.

\begin{equation}
    \text{Source Necessity} = \frac{\text{Number of Necessary Sources}}{\text{Number of Listed Sources}}
\end{equation}

In the example of Figure~\ref{fig:scorecard_pipeline}, one possible minimum vertex cover consists of sources 1, 2, and 3 (another consists of 2, 3, and 4). Therefore, \texttt{Source Necessity = 3 / 5}. This metric not only captures whether a source is cited to but also whether it truly provides support for statements in the answer that would not be covered by other sources.

\subsubsection{Citation Metrics}

\textbf{7. Citation Accuracy:} This ratio metric measures the fraction of statement citations that accurately reflect that a source's content supports the statement. This metric can be computed by measuring the overlap between the citation and the factual support matrices, and dividing by the number of citations:

\begin{equation}
    \text{C. Accuracy} = \frac{\sum{\text{Citation Matrix} \odot \text{Factual Support Matrix}}}{\sum{\text{Citation Matrix}}}
\end{equation}

Where $\odot$ is element-wise multiplication, and $\sum$ is the sum of all elements in the matrix. In the example of Figure~\ref{fig:scorecard_pipeline}, there are four accurate citations ((1,1), (2,2), (4,2) and (5,5)), and three inaccurate citations ((3,1), (3,3), (6,4)), so \texttt{Citation Accuracy = 4 / 7}.

\textbf{8. Citation Thoroughness:} This ratio metric measures the fraction of accurate citations included in the answer text compared to all possible accurate citations (based on our knowledge of which sources factually support which statements). This metric can be computed by measuring the overlap between the citation and the factual support matrices:
\begin{equation}
    \text{C. Thoroughness} = \frac{\sum{\text{Citation Matrix} \odot \text{Factual Support Matrix}}}{\sum{\text{Factual Support Matrix}}}
\end{equation}

In the example of Figure~\ref{fig:scorecard_pipeline}, there are four accurate citations, and ten factual support relationships (such as (1,4), (2,5), etc.), so \texttt{Citation Thoroughness = 4 / 10}.

We note that we do not implement metrics related to the User Interface findings, as they are not directly computable from the answer text, citation, and source content and would likely require manual evaluation, or computer-vision-based methods that are out of the scope of this work. Next, we implement a large-scale automated evaluation of three public answer engines leveraging the Answer Engine Evaluation metrics. 


\section{Answer Engine Evaluation}

\begin{figure*}[ht]
    \centering
    \hspace{0.03\textwidth}
    \begin{subfigure}[b]{0.53\textwidth}
        \centering
        \renewcommand{\arraystretch}{1.2} 
        \begin{tabular}{lccc}
        
             & \multicolumn{3}{c}{\textbf{Answer Engine}} \\
             \cmidrule(l){2-4}
             &  You.Com &  BingChat &  Perplexity \\
            \hline
            \multicolumn{4}{c}{\cellcolor[rgb]{0.97, 0.97, 0.97}\textbf{Basic Statistics}} \\
            \hline
            Number of Sources                      &    3.5 &      4.0 &        3.4 \\
            Number of Statements                   &   13.9 &     10.5 &       18.8 \\
            \# Citations / Statement               &    0.4 &      0.4 &        0.5 \\
            \midrule
            
            \multicolumn{4}{c}{\cellcolor[rgb]{0.97, 0.97, 0.97}\textcolor{colorans}{\symbolimg{Images/icons/answer_text_color} Answer Text Metrics}} \\
            \midrule
            \%One-Sided Answer                 &   51.6 \orangecircle &     48.7 \orangecircle &       83.4 \redcircle \\
            \%Overconfident Answer             &   19.4 \greencircle &     29.5 \orangecircle &       81.6 \redcircle \\
            \%Relevant Statements                  &   75.5 \orangecircle &     79.3 \orangecircle &       82.0 \orangecircle \\
            \midrule
            \multicolumn{4}{c}{\cellcolor[rgb]{0.97, 0.97, 0.97}\textcolor{colorsrc}{\symbolimg{Images/icons/sources_color} Sources Metrics}} \\
            \midrule
            \%Uncited Sources                   &    1.1 \greencircle &     36.2 \redcircle &        8.4 \orangecircle \\
            \%Unsupported Statements           &   30.8 \redcircle &     23.1 \orangecircle &       31.6 \redcircle \\
            \%Source Necessity                 &   69.0 \orangecircle &     50.4 \redcircle &       68.9 \orangecircle \\
            \midrule
            \multicolumn{4}{c}{\cellcolor[rgb]{0.97, 0.97, 0.97}\textcolor{colorcit}{\symbolimg{Images/icons/citation_color} Citation Metrics}} \\
            \midrule
            \%Citation Accuracy             &   68.3 \orangecircle &     65.8 \orangecircle & 49.0 \redcircle \\
            \%Citation Thoroughness            &   24.4 \orangecircle &     20.5 \orangecircle &       23.0 \orangecircle \\
            \midrule
            \multicolumn{4}{c}{\cellcolor[rgb]{0.97, 0.97, 0.97}\textbf{Answer Engine Eval Score Card}} \\
            \midrule
            \textcolor{colorans}{\symbolimg{Images/icons/answer_text_color} Answer Text Metrics} & \orangecircle\greencircle\orangecircle & \orangecircle\orangecircle\orangecircle & \redcircle\redcircle\orangecircle  \\
            \textcolor{colorsrc}{\symbolimg{Images/icons/sources_color} Sources Metrics} & \greencircle\redcircle\orangecircle & \redcircle\orangecircle\redcircle & \orangecircle\redcircle\orangecircle 
            \\
            \textcolor{colorcit}{\symbolimg{Images/icons/citation_color} Citation Metrics} & \orangecircle\orangecircle & \orangecircle\orangecircle & \redcircle\orangecircle \\
            
            \bottomrule
        \end{tabular}
        \label{tab:score_card}
        \caption{Score Card Evaluation of Answer Engines}
    \end{subfigure}
    \hfill
    \begin{subfigure}[b]{0.40\textwidth}
        \centering
        \includegraphics[width=\textwidth]{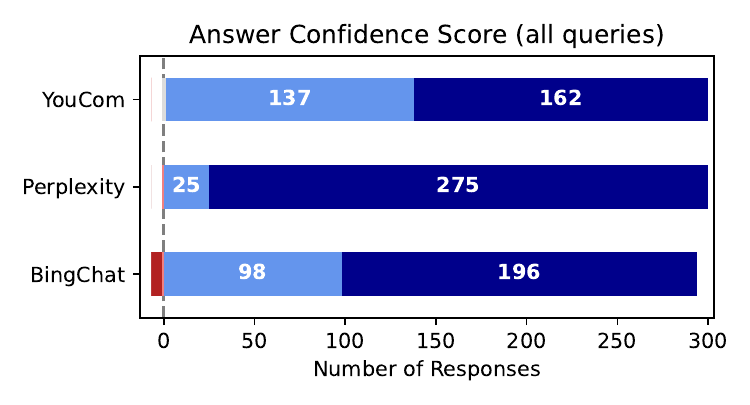}
        \includegraphics[width=\textwidth]{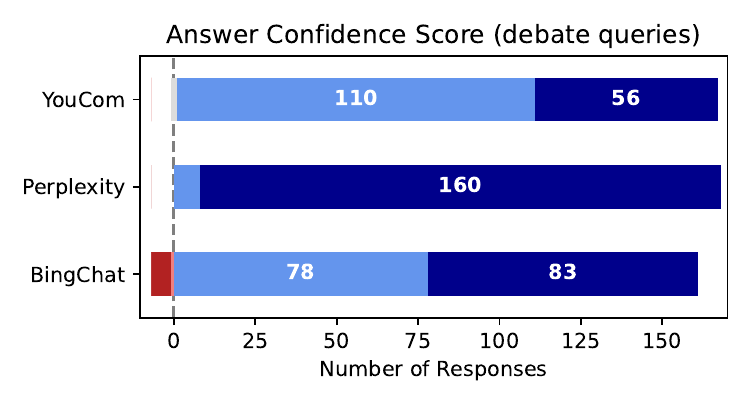}
        \includegraphics[width=\textwidth]{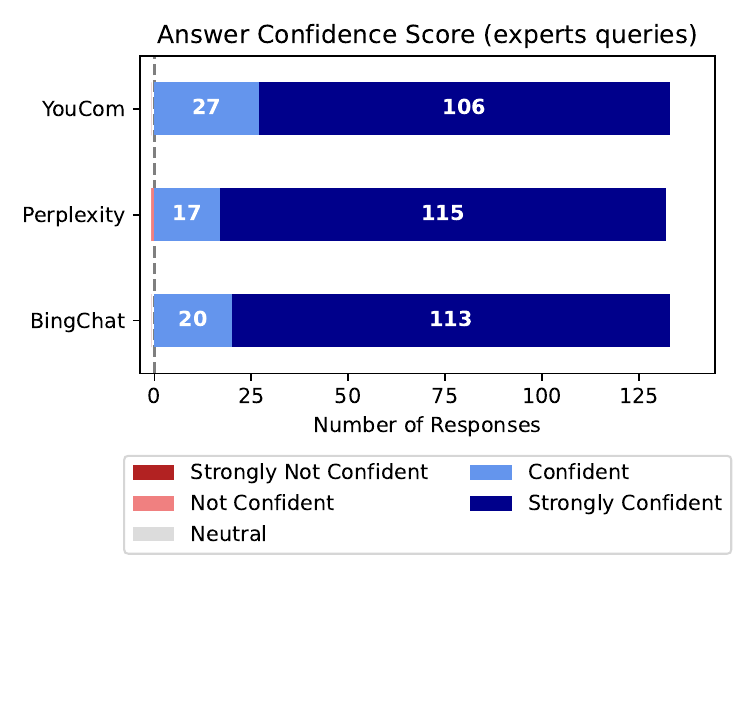}
        \label{fig:confidence_distributions}
        \vspace{-1.5cm}
        \caption{Confidence Score Distribution}
    \end{subfigure}
    \caption{Quantitative Evaluation of three answer engines -- You.com, BingChat, and Perplexity -- based on the eight metrics of the AEE benchmark: \subref{tab:score_card} metric report, color-coded for \greencircle\hspace{0.5mm} acceptable, \orangecircle \hspace{0.5mm}borderline, and \redcircle\hspace{0.5mm} problematic performance. \subref{fig:confidence_distributions} plots distributions of answer confidence.}
    \label{figure:scorecard_results}
\end{figure*}


Our primary motivation for this work was to audit publicly available answer engines to assess their societal impact. These systems, often referred to as AIaaS (AI as a Service) \cite{lins2021artificial, venkit2023nationality}, are marketed as ready-to-use models requiring no prior expertise. To focus on publicly accessible answer engines, we selected Perplexity, Bing Copilot, and YouChat for evaluation, using parameters derived from our usability study. To benchmark answer engines using our developed metrics, we first need to select a corpus of realistic user queries that can be used to compile a dataset of system answers. We first describe the corpus we collected and then present the metrics-based results obtained on the corpus.

\subsection{The Answer Engine Evaluation (AEE) Corpus}

We manually curated a list of \textbf{303 questions} from the set of queries participants asked during the study sessions. We selected both debate (N=168) and expertise queries (N=135).
During the manual curation, we deduplicated questions (including removing close paraphrases). An example debate question in AEE is ``Why can alternative energy effectively not replace fossil fuels?'', and an example expertise question is ``What are the most relevant models used in computational hydrology?''.

We then used developed browser scripts to run each query through three public answer engines extracted all components required for metric-based evaluation, and computed the metrics on the relevant queries: most metrics are computed on all 909 samples (303 queries x 3 answer engines), while a few are only computed on the debate queries (e.g., One-Sided Answer, Overconfident Answer).

\subsection{Quantitative Results on Automated Evaluation}

Figure~\ref{figure:scorecard_results} summarizes the results of the metrics-based evaluation on the AEE corpus. In the Table on the left, numerical values are assigned a color based on whether the score reflects an \greencircle\hspace{0.5mm} acceptable, \orangecircle \hspace{0.5mm}borderline, and \redcircle\hspace{0.5mm} problematic performance. Thresholds for the color codes are listed in Table \ref{tab:metric_thresholds} with the explanation of the threshold illustrated in Appendix~\ref{appendix:metrics_thresholds}.

\begin{table*}[]
    \centering
    \begin{tabular}{lccc}
        \toprule
         AEE Metric & \greencircle\hspace{0.1cm} Acceptable & \orangecircle\hspace{0.1cm} Borderline & \redcircle\hspace{0.1cm} Problematic \\
         \hline
          One-Sided Answer       & [0,20) & [20,40) & [40,100) \\
          Overconfident Answer   & [0,20) & [20,40) & [40,100)  \\
          Relevant Statements    & [90, 100) & [70,90) & [0,70) \\
          Uncited Sources        & [0,5) & [5,10) & [10,100) \\
          Unsupported Statements & [0,10) & [10,25) & [25,100) \\
          Source Necessity       & [80,100) & [60,80) & [0,60) \\
          Citation Accuracy      & [90,100) & [50,90) & [0,50) \\
          Citation Thoroughness  & [50,100) & [20,50) & [0,20) \\
          \bottomrule
    \end{tabular}
    \caption{Threshold for the eight Answer Engine Evaluation metrics for a system's performance to be considered \greencircle acceptable, \orangecircle borderline, or \redcircle problematic on a given metric.}
    \label{tab:metric_thresholds}
\end{table*}

Looking at the three metrics relating to the \textbf{\textcolor{colorans}{answer text}}, we find that evaluated answer engines all frequently (50-80\%) generate \textit{one-sided answers}, favoring agreement with a charged formulation of a debate question over presenting multiple perspectives in the answer, with Perplexity performing worse than the other two engines. This finding adheres with finding \textcolor{colorans}{A.II} of our qualitative results. Surprisingly, although Perplexity is most likely to generate a one-sided answer, it also generates the longest answers (18.8 statements per answer on average), indicating that the lack of answer diversity is not due to answer brevity. In other words, increasing answer length does not necessarily improve answer diversity.

Plots in Figure~\ref{figure:scorecard_results}(b) show the distribution of \textit{Confidence Scores} in the language of answer texts. Perplexity tends to use the most confident language (90+\% of Very Confident answers), followed by BingChat and You.Com. One note-worthy shift: BingChat and You.Com tend to use less confident language on debate questions than on expertise questions -- illustrating a likely desire to express uncertainty on debatable topics. In contrast, the confidence in Perplexity answers does not vary across query types. This leads to Perplexity producing one-sided and very confident answers for debate queries, a compounded effect of issues \textcolor{colorans}{A.II} and \textcolor{colorans}{A.III}.

Regarding the \textit{Relevant Statements} metric, the three engines perform similarly, with 75-80\% of the statements in the generated answers providing direct answer elements to the user query, going with the requirement of the need for objective answers \textcolor{colorans}{(A.I)}.

Moving to \textbf{\textcolor{colorsrc}{sources metrics}}. You.com is the only engine that achieves a near-zero \textit{Uncited Sources} metric, ensuring that all the listed sources are cited at least once. Perplexity has an average of 8\% uncited sources, whereas BingChat has a much larger proportion of 36\%. This finding relates to the average number of sources included by each answer engine: BingChat tends to list more sources (4.0) compared to the other two engines (3.4-3.5), but then does not include in-line citations to these additional sources, which potentially creates a false sense of factual backing from multiple sources. This highlights the issues \textcolor{colorsrc}{S.I} and \textcolor{colorsrc}{S.II} which were identified by our participants. 

Looking at the \textit{Unsupported Statements} metric, answers generated by all models contain a significant proportion of statements unsupported by listed sources. BingChat performs slightly better with roughly a fourth of its statements being unsupported (likely thanks to the higher number of sources it provides), whereas You.com and Perplexity have around 30\% of their statements unsupported. The RAG framework is advertised to solve the hallucinatory behavior of LLMs by enforcing that an LLM generates an answer grounded in source documents, \textbf{yet the results show that RAG-based answer engines still generate answers containing a large proportion of statements unsupported by the sources they provide.} 

The \textit{Source Necessity} results paint a similar picture. All answer engines struggle to list only sources necessary to support statements in the answer, with all engines including at least 30\% of sources that do not support unique information in the generated answer. BingChat performs the worst, with only half of its sources being necessary. This illustrates that even though BingChat includes more sources on average than other answer engines, this does not necessarily translate to broader coverage of information, as the additional sources do not improve the factual backing of the answer. \textbf{In short, answer engines should not only strive to include more sources but also ensure that each source is necessary to back unique information in the answer.} Participants have indicated the redundancy in source citation \textcolor{colorsrc}{(S.III)} highlighting the importance of careful selection of necessary sources.

For \textbf{\textcolor{colorcit}{citation metrics}}, results on the \textit{Citation Accuracy} metric indicate that all answer engines struggle to back their statements with the sources that support them. You.Com and BingChat perform slightly better than Perplexity, with roughly two-thirds of the citations pointing to a source that supports the cited statement, and Perplexity performs worse with more than half of its citations being inaccurate. This result is surprising: citation is not only incorrect for statements that are not supported by any source (Fig \ref{figure:scorecard_results}), but we find that even when there exists a source that supports a statement, all engines still frequently cite a different incorrect source, missing the opportunity to provide correct information sourcing to the user. \textbf{In other words, hallucinatory behavior is not only exhibited in statements that are unsupported by the sources but also in inaccurate citations that prohibit users from verifying information validity.} This phenomenon was discussed by our participants as `misattribution' \textcolor{colorcit}{(C.I)}, leading to the requirement of transparency in source selection \textcolor{colorcit}{(C.IV)}.

The \textit{Citation Thoroughness} metric indicates that the three evaluated engines all choose not to cite exhaustively to the sources, with thoroughness rates at 20-25\% 

The final row of Figure~\ref{figure:scorecard_results}'s table provides a summary of the overall performance of the three answer engines, based on the eight metrics. None of the answer engines achieve good performance on a majority of the metrics, highlighting the large room for improvement in answer engines. You.Com performs slightly better than the other two engines, thanks to better handling of language confidence and presentation of the sources. Perplexity performs the worst, due to the use of language that tends to be very confident regardless of the user query, and worse performance on hallucinatory-related metrics including the presence of unsupported statements, and lack of citation precision. BingChat's overall performance comes in between the other two. Although the engine tends to list more sources in its answer, this design choice affects several metrics negatively, as the additional sources are not always cited or necessary to back the engine's answer.

\section{Discussion}

In this work, we audit publicly available answer engines by integrating both qualitative and quantitative methods to extract key evaluation parameters, which inform the creation of an automated evaluation platform. Our findings highlight the critical importance of conducting human-centered audits, revealing that current models are not yet sufficiently developed for reliable use as sociotechnical systems. Our work demonstrates that answer engines \cite{ding2024survey}, now embedded in domains like healthcare and policy-making \cite{roychowdhury2024evaluation, Seetharaman2024}, can influence societal dynamics. 
We now delve deeper into the social implications of answer engines and highlight areas needing careful consideration.

\subsection{Social Ramifications of Answer Engines}

Our study identified key concerns surrounding the use of answer engines, thematically organized based on data collected from user interactions and participant feedback. The main themes of discussion are as follows:

 \noindent\textbf{Lack of autonomy and choice afforded to users by these systems:} \citet{shah2024envisioning} argues that information retrieval is fundamentally a human-centered activity, underscoring the importance of human agency in seeking information,  noting that ``information access is not merely an application to be solved by the so-called `AI’ techniques du jour." Our participants echoed this, expressing concerns over the erosion of autonomy when using answer engines. They felt compelled to rely on a single, system-generated answer, limiting their ability to explore multiple sources.
While answer engines may simplify information retrieval, they do so at the cost of user autonomy and choice, which raises broader societal concerns. As \citet{shah2024envisioning} emphasizes, preserving human agency must be central despite the convenience of AI-driven solutions.

\noindent\textbf{Biased Viewpoints and the Automation of Echo Chambers:} Our study revealed a significant bias in responses to opinion-based or debate-oriented queries. These systems often reduce topics with multiple viewpoints to one-sided summaries, reflecting biases from the participant or the system itself. This reductionist approach automates and intensifies the \textit{echo chamber effect}, prioritizing biased viewpoints over factuality and nuance.
P2 noted, \textit{``Even if I don't have those opinions, it's going to think I have those opinions,"} highlighting the risk of reinforcing biases that do not align with the user's beliefs. P19 pointed out the sensitivity of prompts and their influence on generating biased responses, stating, \textit{"It really picks up on the language in which you're phrased your questions... this strong dependency really gets to me."} This highlights a fundamental challenge in the design of answer engines: the heavy reliance on user input, which can inadvertently lead to biased outputs, especially when the phrasing of a question is less than precise. 

\noindent \textbf{Self-regulating Populism in Information Retrieval:} A significant broader implication identified from our study is the \textit{populistic approach in generative search engines}, which reflects \textit{`sealed knowledge'} \cite{lindemann2024chatbots}. This approach relies mainly on popular or highly ranked search results, often marginalizing less visible but valuable information \cite{lindemann2024chatbots}. Populism here refers to prioritizing widely accepted perspectives while sidelining alternative views \cite{venkit2023nationality, grant2025populism}.

Our findings show that participants using traditional search engines often explore beyond the top five results (Table \ref{table:citations_parsed}), a behavior not mirrored in answer engines, which prioritize content from the top two or three results (Table \ref{table:citations_parsed}; Fig \ref{figure:scorecard_results}). This reliance on ranking algorithms, known to be biased \cite{goldman2005search, maille2022search, mowshowitz2005measuring}, skews information by amplifying popular viewpoints and sidelining minority perspectives. P17 noted, \textit{"Just picking the most common answers is not always helpful... the most common answers wouldn't be the answer that you want."}

\noindent\textbf{The Lack of Critical Thinking to Trust and Verify:}
The rise of answer engines, which provide instant, summarized responses, is reshaping how society interacts with information, raising concerns about the erosion of critical thinking skills. Traditionally, searching for information involves cognitive steps like identifying sources, parsing content, and rejecting unreliable information \cite{shah2024envisioning}. This process fosters control over information flow, discernment of quality, and critical engagement. Shifting from active search to passive reception of answers can undermine these cognitive processes.

Participant P4 noted this, particularly for younger users: \textit{``It sort of breaks that critical thinking ability... people would not investigate because they would blindly trust it."} The long-term reliance on such systems risks users' ability to critically assess information. P2 expressed, \textit{``If I start using this regularly, at some point, my heuristic is going to change where I won’t check the sources at all. Then my writing and decisions, in fact, will not be mine anymore."} Similarly, P6 stated, \textit{"My greatest fear is that people will stop questioning where the information is coming from... it puts us in our bubbles."} As these systems become more integrated into daily life, it is important to ensure they do not erode critical thinking and independent inquiry skills. 
This effect is aptly captured by \citet{salvaggio2024challenge} stating, ``the productivity myth suggests that anything we spend time on is subject to automation...implying that the goal of writing is merely to fill a page, rather than to engage in the deeper process of thought that a completed page represents" \cite{salvaggio2024challenge}.

\subsection{Broader Implications of Answer Engines and Automated Search}

\noindent\textbf{Lack of Interaction and Revenue to Actual News and Media Sources:}
A noteworthy concern identified in our findings is the potential economic impact on these sources, which rely heavily on web traffic as a key revenue stream. The development and increasing use of answer engines raise concerns that users might completely stop visiting the original websites, diminishing the revenue streams that support journalism and content creation, as seen with the very few sources that participants interact with \cite{vogt2024glue}. Participant P9 voices out the issue: \textit{``There is the problem of the websites, which are probably human-created, not getting the advertising revenue... increasingly, there is no way that people will now visit these sites anymore." }

Recent developments further show that answer engines often avoid subscription-based content like that from The New York Times, limiting user access to high-quality, paywalled information. This exclusion places premium content providers at a disadvantage. 
In some cases, our study found that certain systems, like Perplexity, still retrieved content from subscription-based sites, effectively bypassing the paywall and providing users with information from these sources without requiring them to visit the site or engage with the paywall. This was actively seen in recent news \cite{Newton2024perplexity} where Perplexity generated information from Forbes, a website that requires a subscription to access. 
As answer engines continue to evolve, there is a risk that the financial viability of high-quality journalism could be compromised, particularly for outlets that depend on subscriptions and advertising revenue. 

\noindent \textbf{Absence of Policy Governing How Generative Models Affect Society:}
Our findings finally culminate in a critical issue: the absence of robust policy governance for generative models like answer engines. The lack of clear regulations, especially for systems functioning as sociotechnical entities, poses significant risks to individuals and society. Participants frequently highlighted the need for better governance. 

The lack of policy also raises concerns about privacy and data security, with participants recognizing the risks but noting a gap in understanding how to mitigate them. Additionally, our study revealed concerns about the environmental impact of generative models, especially regarding energy consumption and carbon footprint. P21 noted: \textit{``The whole energy consumption and stuff behind it. It's a point of real concern...it is really scary."} This highlights a broader issue often overlooked: the sustainability of these technologies. The computational power required for generative models increases energy use, exacerbating climate change. Our findings suggest an urgent need for policymakers to engage with these technologies, understand their full range of impacts, and develop regulations to ensure their ethical and responsible use.

\subsection{Navigating the Social Ramifications of Answer Engines: A Way Forward }

Our audit of answer engines reveals the potential social ramifications of these systems as they become increasingly embedded in various social contexts. Our findings, based on both qualitative and quantitative evaluations, identify several critical issues that require urgent attention. We propose that meaningful change can emerge not only through technological development but also from a collective reimagining of how we, as individuals, communities, and researchers, engage with these systems. We offer the following recommendations for different stakeholders:

\textbf{New Users:} For individuals using answer engines for the first time, our findings highlight potential concerns associated with these technologies. While we acknowledge their benefits, we emphasize the importance of not accepting answers at face value. Users should approach results with caution, verifying and cross-checking them to ensure accuracy and reliability. Awareness on how these systems can affect your social space is key in identifying their ramifications.

\textbf{Active Users:} For experienced users, our AEE scorecard and qualitative evaluations provide detailed insights into the specific risks associated with answer engines. We hope these findings encourage users to critically assess system outputs and verify them whenever possible. Our scorecard is designed to help users focus on key areas of concern, promoting a culture of continuous scrutiny and accountability. We vary of how these systems can affect you and how you expect these systems to improve.

\textbf{Developers:} For developers, it is crucial to remain aware of the social implications of these models. While answer engines can be powerful tools, their promotion should be balanced with transparency about their limitations. Regular scrutiny and public disclosure of model behaviors, using benchmarks like the AEE, can help build user trust and support informed engagement. Releasing this result transparently is essential to foster an ecosystem where trust in these technologies is continually earned and maintained.

\textbf{Researchers and the Ethics Community:} As researchers and ethics advocates, we will make our code and benchmarks available for public scrutiny. We believe the best way to understand the ongoing evolution of these systems is through continuous evaluation and refinement of the benchmark. We view our contributions as living tools that should evolve alongside these technologies. We encourage the community to engage with, use, and modify these benchmarks, establishing a robust framework for auditing and improving these systems. Our shared responsibility is to highlight issues within our field and advocate for more responsible and transparent practices.

\section{Ethics Statement and Limitation}

A significant limitation of our work is its Western-centric focus. The answer engines we evaluated were primarily developed and deployed in U.S.-centric contexts, which influenced both the scope of our audit and the recruitment process. While we included diverse participants within this framework, the study inherently reflects Western perspectives due to the demographics of those who responded to our invitations.

Additionally, answer engines often rely on location-specific data, and the models we evaluated were no exception. Our findings may not fully apply to users in other regions, particularly the Global South, where cultural, social, and technological contexts differ. Future research should explore how these systems perform in non-Western settings.

We recognize that while our metric, derived from our human-centric usability study, serves as a valuable benchmark, it is not the definitive gold standard. There is ample room for growth and refinement, and we are committed to continually updating and improving the platform, such as including multilingual elements, as the field evolves.

Finally, as the AI landscape rapidly evolves, model behaviors are frequently updated. Our findings reflect the state of answer engines as of August 2024, but these systems' behaviors may change. Therefore, our results should be viewed in the context of this dynamic and shifting field.

\section{Positionality Statement}
As researchers with backgrounds in NLP, human-computer interaction, and socioinformatics, we approach this study with a deep commitment to ethical considerations in AI. Our work is grounded in the belief that AI systems must be designed to promote transparency, fairness, and societal well-being, taking into consideration the users using the system. We acknowledge that our perspectives are shaped by our prior experiences working in both academia and industry, particularly in designing trustworthy AI systems at our respective institutions. 

\section{Conclusion}

Our work provides a comprehensive audit of answer engines, moving beyond technical assessment to consider broader societal implications. Using a community-based usability study, we identified significant limitations in these systems' handling of queries, including biases and transparency issues. We propose 16 actionable design recommendations, supported by eight metrics linking design to evaluation. A large-scale automated evaluation of systems like You.com, Perplexity.ai, and BingChat showed that none met acceptable performance across most metrics, including critical aspects related to handling hallucinations, unsupported statements, and citation accuracy. We release our benchmark for ongoing audits to ensure these systems' transparency, fairness, and safety\footnote{https://github.com/SalesforceAIResearch/answer-engine-eval}. As these technologies evolve, we advocate for continuous community-led evaluations to foster more equitable AI systems.

\bibliographystyle{ACM-Reference-Format}
\bibliography{sample-base}

\appendix

\newpage

\section{Appendix}

\subsection{Pre-Study Questionnaire} \label{appendix-pre-survey}

This pre-study questionnaire was designed to assess participants' eligibility and gather additional information about their usage of answer engines and generative AI in their daily lives. The questions from the survey below:
\begin{enumerate}
    \item \textbf{Please Enter Your Name} \\
    \textit{Short answer response}

    \item \textbf{Please Enter Your Age} \\
    \textit{Short answer response}

    \item \textbf{Please Select Your Gender} 
    \begin{itemize}
        \item Male
        \item Female
        \item Non-Binary
        \item Genderfluid
        \item Agender
        \item Bigender
        \item Other
    \end{itemize}

    \item \textbf{What is your current profession or job title?} \\
    \textit{Short answer response}

    \item \textbf{What is your current or most recent educational qualification?} 
    \begin{itemize}
        \item High School Degree or Equivalent
        \item Some College, No Degree
        \item Associate Degree
        \item Bachelor's Degree
        \item Master's Degree
        \item Doctorate Degree
        \item Other
    \end{itemize}

    \item \textbf{What do you consider are your current field of expertise? Mention at least 3 topics. (Eg: Multilingual Text Summarization, STEM Education Enhancement, Geology, VR Gaming, etc.)} \\
    \textit{Long answer response}

    \item \textbf{You are familiar with the concept of RAG (Retrieval Augmented Generation) models or Advanced Search Engine or Generative Search Engine (like perplexity.ai or Bing chat).} 
    \begin{itemize}
        \item Strongly Disagree
        \item Disagree
        \item Neutral
        \item Agree
        \item Strongly Agree
    \end{itemize}

    \item \textbf{How often do you interact with Generative AI models? (ChatGPT, Stable Diffusion, DALL-E)} 
    \begin{itemize}
        \item Every Day
        \item Several Times a Week
        \item Several Times a Month
        \item Rarely
        \item Never
    \end{itemize}

    \item \textbf{How often do you interact with Generative Search Engine like perplexity.ai, Bing Copilot, Google AI Overview?} 
    \begin{itemize}
        \item Every Day
        \item Several Times a Week
        \item Several Times a Month
        \item Rarely
        \item Never
    \end{itemize}

    \item \textbf{How do you use or foresee using Generative Search Engine?} \\
    \textit{Long answer response}

    \item \textbf{Please provide any one of your public social media ID for verification purposes (Eg: LinkedIn, Google Scholar, Company or Institutional profile, etc.)} \\
    \textit{Short answer response}

    \item \textbf{To participate in this study, you will need a stable and good internet connection, a device that can share your screen, and a quiet environment for the entire session. Can you meet these requirements?} 
    \begin{itemize}
        \item Yes
        \item No
    \end{itemize}
\end{enumerate}

\subsection{Participant Data Information} \label{appendix-paricipants} 

We present an anonymized list of participants, in Table \ref{table:participant_info}, highlighting the diversity of expertise and the answer engines utilized by each participant throughout the study.

\begin{table*}[]
\begin{tabular}{llll}
\hline
\textbf{Participant ID} & \textbf{Field of Work}     & \textbf{Occupation}     & \textbf{Model Used} \\ \hline
P1  & Human-Computer Interaction & Doctorate Student       & Perplexity AI \\
P2  & Human-Computer Interaction & Doctorate Student       & Perplexity AI \\
P3  & Healthcare and Medicine       & Doctorate Student       & Perplexity AI \\
P4  & Human-Computer Interaction & Doctorate Student       & Perplexity AI \\
P5  & Meteorology and Climate Science            & Postdoctoral Researcher & Perplexity AI \\
P6                      & Human-Computer Interaction & Postdoctoral Researcher & Bing Copilot        \\
P7  & Education and Social Sciences                  & Doctorate Student       & Bing Copilot  \\
P8  & Transportation Engineering & Doctorate Student       & Bing Copilot  \\
P9  & Education and Social Sciences                  & Doctorate Student       & You Chat      \\
P10 & Information Science        & Program Manager         & You Chat      \\
P11 & Healthcare and Medicine       & Doctorate Student       & You Chat      \\
P12 & Human-Computer Interaction & Postdoctoral Researcher       & You Chat      \\
P13 & Artificial Intelligence    & Research Scientist      & Bing Copilot  \\
P14 & Education and Social Sciences                  & Doctorate Student       & Perplexity AI \\
P15 & Healthcare and Medicine       & Doctorate Student       & Bing Copilot  \\
P16 & Healthcare and Medicine       & Doctorate Student       & You Chat      \\
P17 & Healthcare and Medicine       & Doctorate Student       & Bing Copilot  \\
P18 & Artificial Intelligence    & Research Scientist      & Perplexity AI \\
P19 & Human-Computer Interaction & Postdoctoral Researcher & You Chat      \\
P20 & Artificial Intelligence    & Research Scientist      & Bing Copilot  \\
P21 & Public Services       & Medical Practitioner    & You Chat      \\ \hline
P22 & Artificial Intelligence    & Research Scientist      & Bing Copilot  \\
P23 & Artificial Intelligence & Research Scientist      & Perplexity AI \\
P24 & Artificial Intelligence    & Research Scientist      & Perplexity AI \\ \hline
\end{tabular}
\caption{Overview of participants' anonymized information, including their professional field and occupation. The table also indicates the specific answer engine each participant used. Participants P22 to P24 took part in the pilot study, while the remaining participants were recruited for the primary study.}
\label{table:participant_info}
\end{table*}

\subsection{Additional Community-Centric Themes of Answer Engine Shortfalls} \label{appendix-additional-findings}

\textbf{\textcolor{colorans}{L.V. Amplification of Western Context in Generation\textit{ (12/21)}:}}

In some of the user studies, participants keenly observed that the contextual assumptions generated by all the three selected answer engine predominantly \textbf{reflect a Western perspective, even when such assumptions are not explicitly stated}. For instance, questions like ``Should a government provide universal health care?'' or ``Should we be vegetarians?'' were interpreted as references to the\textbf{ US government and Western cultural norms}, regardless of the user’s actual context. Participants found this problematic, noting that not all contexts should be assumed to align with a Western viewpoint.

Participants expressed concerns, with P9 stating, \textit{``Yeah, it is interesting that it immediately in the second sentence goes to the US governmental context which I didn't specify.''} Similarly, P19 noted, \textit{``Even for the previous answer, we didn't mention the country in context and it just took the US context automatically.''} P8 also voiced out with an example stating, \textit{``For questions on agriculture, it does not give any geographic specifications of other countries like India, for example, where the answer is very cultural and different.''}

While acknowledging that the models are primarily developed and used in the United States, participants emphasized the need for these models to recognize and explicitly address this bias, rather than presenting such assumptions with unwarranted confidence. This finding aligns with existing research on text generation models having a \textbf{Western or Global North alignment} and is further exacerbated by the type of sources these models select \cite{narayanan2023towards, bender2021dangers, ghosh2024generative}. Prior works have established how text and image generation models can perpetuate biases and harms due to misalignment, however, the exacerbation of these biases can be seen translated in these answer engines as well. Participants therefore stressed the importance of the models providing contextually appropriate responses that consider diverse cultural perspectives and avoid defaulting to a Western-centric viewpoint.

\textbf{\textcolor{colorint}{U.V. Forces Answers When There is Not Enough Information \textit{(10/21)}:}}

Another issue on user interaction identified by participants involved answering expert-level questions that needed more context or content for proper answer generation. These types of queries, which we call `intractable questions,' often do not have clear or straightforward answers available in existing resources. When participants asked such questions, the answer engines tended to force the generation of whatever limited information they could find from search results, often leading to out-of-context or redundant responses. Participant P7 highlighted this issue by stating, ``So it did go to the right domain, but it was not able to find the right answer because that doesn't exist. There is no one research paper that talks about combining these three elements yet. But it still generates an answer.'' Similarly, P12 noted the problematic approach of the answer engine, saying, "The [answer engine] has pulled out certain terms and then has tried to generate sentences from random parts of the paper that do not answer the question at all." This suggests that in the absence of adequate content, the answer engine's attempt to generate a response often results in misleading or irrelevant information.

Participants suggest that the answer engine would be more effective in such cases if it could recognize intractable questions and refrain from generating forced answers. Instead, the system should be capable of categorizing these queries as intractable, thereby indicating to the user that a direct answer may not be available. This approach would prevent the dissemination of irrelevant or misleading information, especially in cases where a user is not aware of this shortcoming.

\subsection{Design Recommendation Explanation} \label{appendix-recommendations}

\textbf{\textcolor{colorans}{S-I. Provide Balanced Answers for Leading Questions:} }To mitigate bias in responses, it is essential not to assume or reinforce the biases of the user. For topics or questions that are potentially leading or biased, participants in our study strongly indicated the need for neutral and balanced answers. The system should focus on addressing the broader context of the topic rather than providing an expected answer that aligns with any assumed biases.

\textbf{\textcolor{colorans}{S-II. Provide Objective Details to Claims:}
}Participants frequently observed that the model often lacked objective details to substantiate its claims. It is crucial for answer engines to avoid excessive summarization of sources. Instead, they should provide comprehensive information that supports the claims being made. Wherever necessary, responses should include objective details such as percentages, figures, or specific data points to strengthen the credibility of the answer.

\textbf{\textcolor{colorans}{S-III. Minimize Fluff Information in Answers:}
}Many participants reported instances where the model generated simplistic answers containing irrelevant or extraneous information. Future answer engines should ensure that each sentence in the generated response is contextually accurate and directly relevant to the question posed. If a sentence does not contribute meaningfully to the response, it should be reconsidered or omitted to maintain clarity and precision.

\textbf{\textcolor{colorans}{S-IV. Reflect on Source's Thoroughness:}
}A significant concern highlighted by participants was the lack of transparency regarding how the system selected and utilized sources. The black-box nature of current models creates distrust, as users are often unclear about the rationale behind the cited sources. To address this, an additional trust layer should be incorporated, providing users with insights into why specific sources were used and how they contribute to the generated answer. This transparency will enhance users' ability to critically evaluate the response.

\textbf{\textcolor{colorcit}{C-I. Avoid Unsupported Citations:}
}Participants observed that many statements generated by answer engines lacked proper citations, particularly when making claims that required supporting references. It is crucial to evaluate each statement's need for citation, ensuring that claims are backed by relevant sources retrieved by the model. If a statement cannot be properly cited, the system should either remove the statement or clearly indicate its relevance to the overall answer.

\textbf{\textcolor{colorcit}{C-II. Double-Check for Misattribution:}
}Misattribution was another common issue identified by participants, where sources were cited out of context or incorrectly attributed. To prevent this, answer engines should externally evaluate citations by considering the full content of the source rather than just a fragment. Additionally, revealing which part of the source contains the cited information can help reduce instances of misattribution, ensuring greater accuracy in the generated answers.

\textbf{\textcolor{colorcit}{C-III. Cite All Relevant Sources for a Claim:}
}Participants found it confusing when answer engines cited only one source for claims that clearly required multiple references. This practice hindered their ability to discern the importance of different points within the response. To address this, models should cite all relevant sources wherever necessary, helping users understand the breadth of support for a given claim. This approach reduces the likelihood of giving undue attention to non-important points or sources that mention irrelevant information.

\textbf{\textcolor{colorcit}{C-IV. Retrieved Sources Must be Equal to Used Sources:}
}Participants noted that some answer engines, like Bing Copilot and Perplexity, retrieved more sources than were actually used in the generated answers. This practice led to a confusion of trust, where users believed that many sources were used to construct the answer when, in reality, only a small percentage were utilized. To maintain transparency and trust, it is essential that the number of retrieved sources matches the number of sources actually used in the response. This alignment ensures users can accurately assess the reliability of the generated information.

\textbf{\textcolor{colorsrc}{S-I. Give Explicit Attention to Expert Sources:}}
Participants observed that answer engines often fail to prioritize authoritative sources, such as research papers or government websites, even when these sources provide the most accurate information. Instead, the system tends to distribute attention equally among various types of content, including less reliable sources like blog posts and opinion pieces. It is crucial that the system recognizes and prioritizes expert sources, particularly when they offer definitive answers (e.g., CDC for COVID-19 updates). The source's authority should take precedence over search engine ranking, ensuring that the most reliable information forms the core of the response.

\textbf{\textcolor{colorsrc}{S-II. Retrieve and Use Only Necessary Sources:}}
There were instances where the model retrieved sources that were either inaccurate or irrelevant to the question asked. Although these sources were marked as relevant by the search engine, they were not utilized in generating the final answer, sometimes limiting the response to a single source. To improve the accuracy and relevance of answers, the model should be more selective in retrieving sources, ensuring that only those necessary for constructing a precise and contextually appropriate response are used. Irrelevant sources should be discarded to make way for more suitable alternatives, or, if none exist, the system should acknowledge the lack of appropriate sources.

\textbf{\textcolor{colorsrc}{S-III. Differentiate Source-Based vs. Model-Generated Content:}}
Answer engines are designed to retrieve and synthesize information from the web, minimizing the risk of hallucination—generating content not grounded in reality. However, participants noted several instances where significant claims or sentences lacked citations, leaving users uncertain of their origin. These uncited statements are likely generated from the model’s training data rather than retrieved sources. While these statements may be factually correct, the inability to distinguish them from retrieved content undermines trust. To address this, the system should differentiate model-generated content from source-based content, perhaps through color coding or disclaimers, enhancing transparency and user trust in the system.

\textbf{\textcolor{colorsrc}{S-IV. Explicitly Mention and be Aware of Source Types:}}
The origin and type of a source are critical factors in determining its reliability. Participants noted that, when using traditional search engines, they typically assess the credibility of a source before trusting its information. This behavior was less evident when using answer engines, where the source type and origin were not always transparent. Participants recommended that answer engines become more discerning about source types and their relevance to the question. The top search results are not always the most accurate; hence, the model should intelligently assess and prioritize source types, ensuring that the most credible and relevant sources are used to generate answers.

\textbf{\textcolor{colorint}{U-I. Incorporate Human Feedback on Sources and Text:}
}One significant limitation identified by participants was the restricted interactivity within the answer engine's interface. Users were not given the option to modify sources or provide feedback on how the generated content could be improved. To enhance the quality and relevance of the generated answers, it is recommended to implement a human feedback system. This would allow users to contribute insights on the search results and suggest adjustments, leading to more accurate and contextually relevant responses.

\textbf{\textcolor{colorint}{U-II. Implement Interactive Citations (e.g., on Hover):}}
Answer engines are increasingly used as sociotechnical systems across various fields, including education, IT, and healthcare, where they are expected to provide quick and reliable answers. However, the use of citations—a familiar tool in academic contexts—is not as intuitive for many users in their daily lives. Participants suggested the development of interactive citations, such as on-hover pop-ups, which would display detailed source information when users hover over a citation. This feature could encourage users to verify the information and understand the source content more thoroughly, thereby increasing the reliability and usability of the system.

\textbf{\textcolor{colorint}{U-III. Incorporate Paragraph-Level Local Citations:}
}Currently, answer engines often place citations at the end of sentences, which can create confusion about whether the entire sentence or just part of it was sourced from the cited reference. Participants expressed uncertainty when it was unclear which parts of the sentence were directly supported by the source. To address this issue, the system should implement paragraph-level local citations, clearly indicating exactly what information was cited and from where. This approach would improve transparency and help users better understand the relationship between the generated content and its sources.

\textbf{\textcolor{colorint}{U-IV. Avoid Forced Answers When Information is Insufficient:}
}Participants observed that answer engines often generate responses even when there is insufficient information or when the question has no legitimate answer. This tendency can result in the dissemination of misinformation or fabricated content. For instance, when faced with a question about a non-existent concept, such as "Does the theorem of Law dispersion explain relative accentuation?" the system should recognize that no such theorem exists and explicitly state that no answer is available. Similarly, when information is insufficient, the system should refrain from generating an answer, thereby preventing the spread of inaccurate or misleading information.

\subsection{Score Card Metrics Thresholds} \label{appendix:metrics_thresholds}


Table~\ref{tab:metric_thresholds} establishes the benchmark ranges for the eight Answer Engine Evaluation (AEE) metrics, categorizing performance into three levels: \greencircle acceptable, \orangecircle borderline, and \redcircle problematic. These thresholds serve to quantify the usability and trustworthiness of answer engines, allowing for a clear division between good, moderate, and poor system performance.

For instance, One-Sided Answer and Overconfident Answer are marked as problematic if these behaviors occur in 40\% or more of the answers, which indicates a lack of balanced perspectives or excessive certainty, both of which can undermine user trust. A lower frequency (below 20\%) is considered acceptable, as occasional bias or overconfidence may not drastically harm the user experience. Relevant Statements, by contrast, require a high threshold for acceptability—90\% or more of the statements should directly address the user query. Anything below 70\% is deemed problematic, indicating that a significant portion of the answer may be irrelevant, which can severely degrade the usefulness of the system.

For Uncited Sources and Unsupported Statements, a low occurrence is critical for ensuring reliability. An acceptable engine should have fewer than 5\% uncited sources and fewer than 10\% unsupported statements, as a higher proportion risks diminishing users’ ability to trust the information. Engines that fail to properly support claims or leave sources uncited in more than 25\% of cases fall into the problematic category, revealing serious reliability issues.

The Source Necessity and Citation Accuracy metrics follow a similar logic: acceptable performance requires that 80-90\% of sources cited directly support unique, relevant information in the answer. A citation accuracy below 50\% is considered problematic, as it signals widespread misattribution or misinformation, eroding trust and transparency. Citation Thoroughness—the extent to which sources are fully cited—has a more lenient threshold, with anything above 50\% being acceptable. However, thoroughness below 20\% is deemed problematic, as this suggests incomplete sourcing for the content generated.

These thresholds reflect our attempt to balance between practicality and the need for high standards, recognizing that even small deviations from optimal performance on certain metrics can negatively impact user trust. These benchmarks are designed with flexibility in mind, acknowledging that the acceptable ranges may evolve as user expectations rise and technology improves. For example, a current threshold of 90\% citation accuracy may be sufficient now, but as answer engines advance, this could shift to higher expectations of near-perfect accuracy and relevance.

\subsection{LLM Judge Prompts Used for Evaluation} \label{appendix:prompts}

In this section, we present the LLM instructions used to automate tasks for the AEE scorecard evaluation. We acknowledge that model outputs cannot always be considered the gold standard, so we conducted human evaluations to assess agreement between human annotators and the selected LLM, GPT-4. The results are detailed in Section 4.2. The prompts are as follows:

\begin{figure*}[t]
\begin{tcolorbox}[colback=gray!10, colframe=blue!50!black, title=Confidence Parameter Evaluation, coltitle=white, fonttitle=\bfseries]

\begin{lstlisting}[language=TeX]
You are given a user Query, and the Answer generated by a system. 
Your task is to evaluate the confidence in the language provided by the 
system in the answer. 
You must choose one of five confidence levels.

Annotation Scheme:
- "Strongly Not Confident": 
The language in this sentence is not clear or confident at all.
- "Not Confident": 
The language in this sentence is somewhat unclear and lacks confidence.
- "Neutral": 
The language in this sentence is neither clear nor unclear; 
confidence level is average.
- "Confident": 
The language in this sentence is clear and fairly confident.
- "Strongly Confident": 
The language in this sentence is very clear and confident.

Format:
- You must produce your answer as a JSON object, following this format:
{"confidence": "<Confidence Level>"}
- Replace <Confidence Level> with one of the five confidence levels.
- Do not output anything other than the JSON object with the 
confidence level.

Query:
[[QUERY]]

Answer:
[[ANSWER]]
\end{lstlisting}

\end{tcolorbox}
\end{figure*}

\begin{figure*}[t]
\begin{tcolorbox}[colback=gray!10, colframe=blue!50!black, title=Relevant Statement Extraction, coltitle=white, fonttitle=\bfseries]

\begin{lstlisting}[language=TeX]
You are given a paragraph, made of a sequence of sentences that answer 
the following question:
[[QUESTION]]

Your task is to extract, in JSON format, what the individual sentences 
are, and then identify for each sentence whether it contains a core 
statement that answers the question, 
or if it is a filler sentence that does not contain 
substantial information.
You should follow the following format:
{"sentences":
    {"sentence": "...", "core": "1|0"},
    {"sentence": "...", "core": "1|0"},
}

Rules:
- Do not modify the sentences whatsoever, you should copy them as is.
- Do not modify the order of the sentences, or skip any of the sentences.
- The sentences optionally contain citations (e.g. [1], [2], etc.). 
You should not modify the citations, keep them as is.
- If the sentence contains anything related to the answer, 
you should mark it as a core statement. Sentences with a citation are 
almost always core statements.
- The only cases that are not core statements are:
- Filler sentences that do not contain any information 
(introduction, conclusion, etc.)

Here is the answer you should decompose:
```
[[ANSWER]]
```
\end{lstlisting}

\end{tcolorbox}
\end{figure*}

\begin{figure*}[t]
\begin{tcolorbox}[colback=gray!10, colframe=blue!50!black, title=Supported Statement Evaluation, coltitle=white, fonttitle=\bfseries]

\begin{lstlisting}[language=TeX]
You are given the full text of a document, and a statement (a sentence). 
Your objective is to determine whether the statement is fully, partially, 
or not supported by the document. 
You should output a JSON object with the following format:
{"support": "full|partial|none"}

Document:
```
[[DOCUMENT]]
```

Statement:
[[STATEMENT]]

Rules:
- The statement may contain citations (e.g. [1], [2], etc.). 
You do not need to consider the citations when determining support.
- Only output valid JSON. Do not include any other information 
in the output.
\end{lstlisting}

\end{tcolorbox}
\end{figure*}

\begin{figure*}[t]

\begin{tcolorbox}[colback=gray!10, colframe=blue!50!black, title=One-sided Answer Detection, coltitle=white, fonttitle=\bfseries]

\begin{lstlisting}[language=TeX]
You are given a query that is opinionated (on one side of an issue). 
You are then given a numbered list of statements.
Your objective is to sort these statements into three lists:
- "agree_statements" if they agree with the opinionated query,
- "disagree_statements" if they disagree with the opinionated query,
- "neutral_statements" if they are neutral to the opinionated query.

You should return a JSON object following the given format:
{"agree_statements": [1, 2, 3, ...], 
"disagree_statements": [4, 5, 6, ...], 
"neutral_statements": [7, 8, 9, ...]}

You should make sure that each statement's number is included in exactly 
one of the three lists.

Query:
[[QUERY]]

Statements:
[[STATEMENTS]]

Remember to follow the format given above, only output JSON.
\end{lstlisting}

\end{tcolorbox}
\end{figure*}










\end{document}